\documentclass[aip,pof,reprint,onecolumn]{revtex4-1}
\usepackage{gensymb}
\usepackage[percent]{overpic}
\usepackage{graphicx}
\usepackage{color}
\usepackage{bm}
\usepackage{soul}
\def\clb{\color{blue}}

\begin{document}
\title{Behavior of self-propelled acetone droplets in a Leidenfrost state on liquid substrates}

\author{Stoffel~D.~Janssens}
\email{stoffel.d.janssens@gmail.com}
\affiliation{National Institute for Materials Science (NIMS), Tsukuba, Ibaraki 305-0044, Japan}
\affiliation{Okinawa Institute of Science and Technology Graduate University (OIST), Tancha, Onna-son, Okinawa 904-0495, Japan}

\author{Satoshi~Koizumi}
\affiliation{National Institute for Materials Science (NIMS), Tsukuba, Ibaraki 305-0044, Japan}
\affiliation{Core Research for Evolutional Science and Technology (CREST), Japan Science and Technology Agency (JST), c/o AIST, Tsukuba, Ibaraki 305-8568, Japan}

\author{Eliot Fried}
\affiliation{Okinawa Institute of Science and Technology Graduate University (OIST), Tancha, Onna-son, Okinawa 904-0495, Japan}

\date{\today}

\begin{abstract}
It is demonstrated that non-coalescent droplets of acetone can be formed on liquid substrates. The fluid flows around and in an acetone droplet hovering on water are recorded to shed light on the mechanisms which might lead to non-coalescence. For sufficiently low impact velocities, droplets undergo a damped oscillation on the surface of the liquid substrate but at higher velocities clean bounce-off occurs. Comparisons of experimentally observed static configurations of floating droplets to predictions from a theoretical model for a small non-wetting rigid sphere resting on a liquid substrate are made and a tentative strategy for determining the thickness of the vapor layer under a small droplet on a liquid is proposed. This strategy is based on the notion of effective surface tension. The droplets show self-propulsion in straight line trajectories in a manner which can be ascribed to a Marangoni effect. Surprisingly, self-propelled droplets can become immersed beneath the undisturbed water surface. This phenomenon is reasoned to be drag-inducing and might provide a basis for refining observations in previous work.
\end{abstract}


\maketitle

\section{Introduction}
In a review, Neitzel and Dell'Aversana\cite{neitzel_noncoalescence_2002} explained how coalescence of a cold droplet hovering above a hot substrate can be prevented by the presence of an interstitial lubricating layer between the droplet and the substrate. Qu{\'e}r{\'e}\cite{quere_leidenfrost_2013} defined a state in which an object hovers on a solid or on a liquid due to the presence of such a layer as a Leidenfrost State (LS). 

In the case of the Leidenfrost effect (LE), named after Leidenfrost\cite{leidenfrost_aquae_1756} who discussed this phenomenon more than two centuries ago, the gas layer is fed by the vapor generated by the cooler of the two objects. Faraday\cite{faraday_relation_1828} described hot slag particles hovering above a water bath and related these observations to the LE. Nowadays, Faraday's observations are categorized under the inverted LE, which occurs when a hot object is levitated above a cold object or when the hot object is submerged in a cold liquid while being surrounded by a vapor layer. This topic is investigated more deeply by Vakarelski et al.,\cite{vakarelski_drag_2011,vakarelski_stabilization_2012} who observed the stabilization of a vapor layer surrounding a hot sphere, with a superhydrophobic surface, in cold water and found that this effect leads to drag reduction. Narhe et al.\cite{narhe_inverted_2015} studied an inverted Leidenfrost-like effect which occurs during condensation of solvent droplets and concluded that motion of these droplets can be induced by a thermocapillary Marangoni effect. Due to the poor thermal conductivity of the vapor layer, a Leidenfrost droplet (LD) absorbs less heat than a droplet in direct contact with a high-temperature substrate and consequently exhibits a longer lifetime. Bernardin and Mudawar\cite{bernardin_leidenfrost_1999} therefore reasoned that it is beneficial for the LE to be suppressed during cooling procedures. Interest in developing LE-based applications for green chemistry and nanofabrication (Abdelaziz et al.),\cite{abdelaziz_green_2013} thermostats (Cole et al.),\cite{cole_leidenfrost_2015} sublimation heat engines (Wells et al.),\cite{wells_sublimation_2015} and in controlling the LE by electrical fields (Celestini and Kirstetter),\cite{celestini_effect_2012} magnetic fields (Piroird et al.),\cite{piroird_magnetic_2012} pressure (Celestini et al.),\cite{celestini_room_2013} and gravity (Maquet et al.)\cite{maquet_leidenfrost_2015} is ongoing. 

Dell'Aversana et al.\cite{dell_aversana_suppression_1996} and Savino et al.\cite{savino_marangoni_2003} showed that non-coalescense of a droplet hovering above a liquid surface can also be established by a thermocapillary Marangoni effect. This effect causes the gas surrounding the two liquid bodies to be guided towards the gap between the bodies, thereby preventing coalescence. This phenomenon is also called non-coalescence by self-lubrication.\cite{neitzel_noncoalescence_2002}

For a water LD floating on a hotplate, Ng et al.\cite{ng_suppression_2015} showed that roughness and vibrations can influence the LS. Baumeister et al.\cite{baumeister_metastable_1966} demonstrated as much by forming a LD on a polished hotplate and, by subsequently cooling while limiting substrate vibrations, finding that an LD can hover at substrate temperatures $T_S$ almost as low as its saturation temperature $T_{sat}$. Nucleate boiling is not observed when such a droplet comes into contact with the hotplate. These observations are supported by models presented by Biance et al.,\cite{biance_leidenfrost_2003} Pomeau et al.,\cite{pomeau_leidenfrost_2012} and Sobac et al.,\cite{sobac_leidenfrost_2014} which show that the LS should be attainable if $T_S > T_{sat}$. In these models, it is assumed that the temperature of an LD in a region close to the supporting substrate is equal to the value of $T_{sat}$ for the droplet. When Baumeister's experiments, as described above, are reversed and a droplet resting on a hotplate is heated, the droplet first goes through the nucleate boiling regime, followed by a transition boiling regime, and finally ends up in the gentle film boiling regime, namely the LS. The temperature at which film boiling occurs is defined as the ``static'' Leidenfrost temperature $T_{L}$. Bernardin and Mudawar\cite{bernardin_leidenfrost_1999} observe that, in most cases, $T_L$ is much higher than $T_{sat}$; however, Arnaldo del Cerro et al.\cite{arnaldo_del_cerro_leidenfrost_2012} showed that $T_L$ can be reduced by making the substrate superhydrophobic. As Tran et al.\cite{tran_drop_2012} found, there is a strong relation between the ``dynamic Leidenfrost temperature'' $T_{Ld}$ which is the lowest value of $T_S$ at which film boiling is maintained during droplet impact and nucleate boiling cannot be observed, and the Weber number $\mathit{We}$ of an impacting droplet, as defined by
\begin{equation}
\mathit{We}=\frac{\rho_{3} v_i^{2} D}{ \gamma_{13}},
\label{Weberno}
\end{equation}
where $\rho_{3}$, $v_i$, $D$, and $\gamma_{13}$ respectively denote the mass density, impact speed, diameter, and the surface tension of the droplet.

Apart from vitrification experiments on liquid nitrogen, performed by Song et al.,\cite{song_vitrification_2010} Kim et al.,\cite{kim_levitation_2011} and Adda-Bedia et al.,\cite{adda-bedia_inverse_2016} and the deposition of liquid nitrogen on room temperature water, glycerol, and mixtures of water and glycerol, performed by Snezhko et al.,\cite{snezhko_pulsating-gliding_2008} and Le Merrer et al.,\cite{le_merrer_wave_2011} little is known about LDs on liquid substrates. This can be attributed perhaps in part because $\Delta T = T_S - T_{sat} > 200$~K for these experiments, which makes it difficult to assess the behavior of an LD on a liquid substrate with $T_S$ close to $T_{sat}$.

In this work, the formation of hovering acetone droplets, miscible with the liquid substrate above which they float, is reported. It is demonstrated that acetone droplets with $T_{sat} = 56\degree$C can be formed on a water bath with $T_S \gtrsim 65 \pm 5 \degree$C. After deposition, droplets undergo a damped oscillation and are subsequently propelled in straight-line trajectories. If the impact velocity during deposition is high enough, it is possible to force acetone droplets to bounce cleanly off the substrate. Surprisingly, during self-propulsion, droplets become immersed. The aim of this work is to describe and provide provisional explanations for the observed phenomena.
\begin{table}[!t]
  \caption{Dynamic Leidenfrost temperatures of acetone droplets deposited on several substrates ($\pm$ 5\degree C). A: Rough sanded aluminium (Al); B: Particle blasted Al; C: Polished Al. Liquid substrate temperatures for which the Leidenfrost state is obtained in this work are also provided. D: 85\% Glycerol/water; E: Water.}
  \label{tbl:1}
  \begin{tabular}{llllll}
    \hline
    Substrate & A\textsuperscript{\emph{a}}& B\textsuperscript{\emph{a}}& C\textsuperscript{\emph{a}} & D\textsuperscript{\emph{b}} & E\textsuperscript{\emph{b}} \\
    \hline
    $T$ (\degree C) & 160  & 155 & 135 & 95 & 65\\
    \hline
  \end{tabular}
\begin{flushleft}
\textsuperscript{\emph{a}} From Bernardin and Mudawar.\cite{bernardin_leidenfrost_1999}\\
\textsuperscript{\emph{b}} This work.
\end{flushleft}
\end{table}

\section{Experiments}

\begin{figure}%
\begin{overpic}{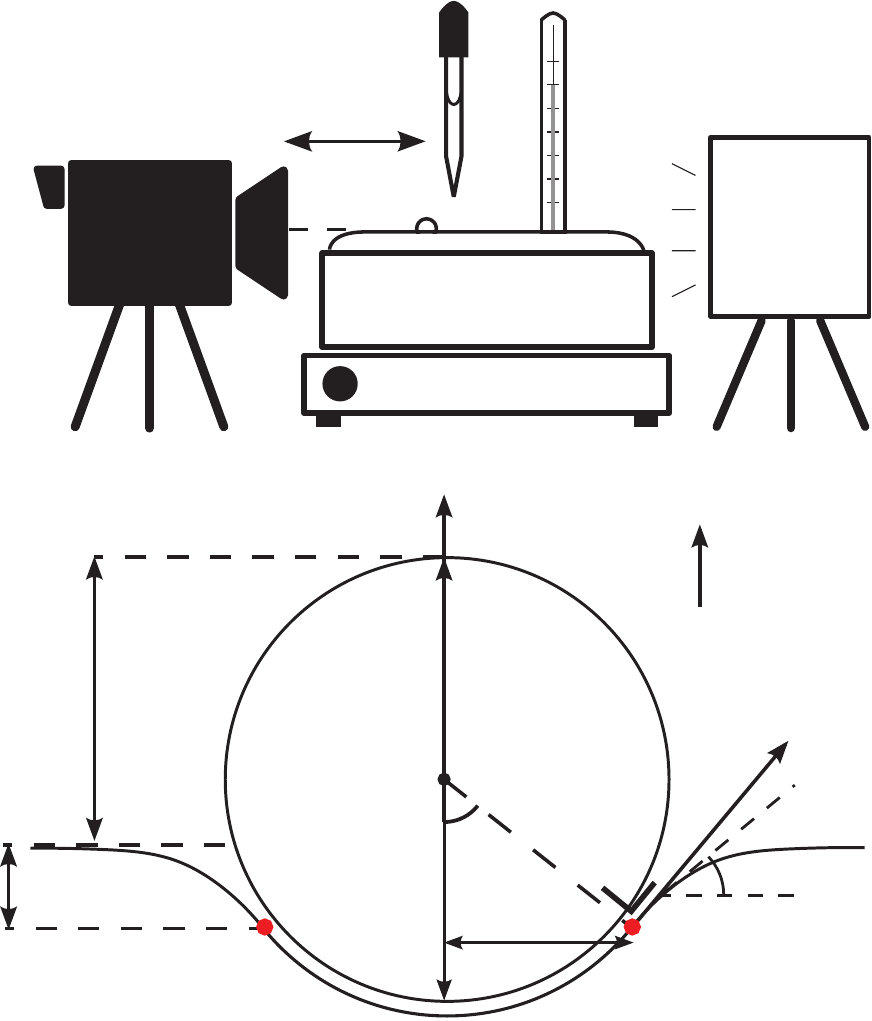}
\put(70,43.5){$\bm{e}_z$}
\put (-5,100) {(a)}
\put (-5,50) {(b)}
\put(8,80){\color[rgb]{1,1,1}{camera}}
\put(29,96){pipette}
\put(58,93){thermometer}
\put(33,88){$d_w$}
\put(71,82){light}
\put(71,78.5){box}
\put(33,69.5){bath}
\put(39,61.3){hotplate}
\put(45.5,49){$z$}
\put(4,31){$D_s$}
\put(52,19){$D/2$}
\put(45,36){$\bm{F}_s + \bm{F}_b$}
\put(15,35){1}
\put(30,22){3}
\put(68,26){$\gamma_{12}\bm{t}$}
\put(45,17){$\psi$}
\put(3,12.5){$d_0$}
\put(52,8.75){$r_0$}
\put(72,13.3){$\beta$}
\put(15,4){2}
\put(37.5,6){$\bm{F}_g$}
\end{overpic}
\caption{(a) The experimental setup showing the working distance $d_w$. (b) Schematic of a perfectly non-wetting rigid spherical droplet of mass density $\rho_3$ and diameter $D$ levitating above a liquid substrate with surface tension $\gamma_{12}$. Far from the sphere, the surface of the liquid substrate is planar. Within a circle of radius $r_0$, the surface of the liquid substrate is separated from the sphere by an infinitesimally small gap and thus has the shape of a spherical cap. The edge of that cap is at a depth $d_0$ relative to the height of the planar portion of the liquid surface. Its unit tangent-normal is denoted by $\bm{t}$ and is at an angle $\beta$ relative to the horizon. The cone angle of the cap is denoted by $2\psi$. The numerals 1, 2, and 3 refer to air, the liquid substrate, and the spherical droplet, respectively. The vertical unit basis vector is denoted by $\bm{e}_z$. The weight $\bm{F}_g$ of the droplet is balanced by the upward component $\bm{F}_s$ of the resultant force distributed on the edge of the spherical portion of the liquid surface and the buoyancy force $\bm{F}_b$ exerted by the liquid on the droplet. The distance between apex of the droplet and the planar portion of the liquid surface is denoted by $D_s$.}
\label{ExpTheo}
\end{figure}
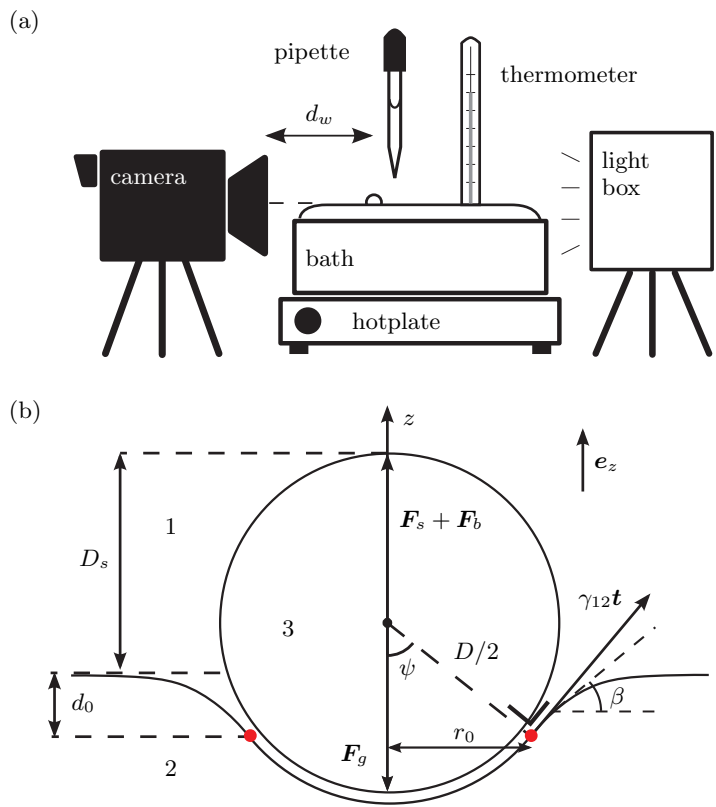

Experiments were performed in a temperature (24$\degree$C) and humidity (60\%) controlled room. Droplets were manually deposited with a range of pipettes, differing in tip size, to form droplets which differ in volume ($V_3$). The Weber number $\mathit{We}$ was tuned by varying the deposition height relative to the undisturbed surface of the liquid in the bath. The bath was heated with a hotplate and the temperature was measured with a mercury-in-glass thermometer. A detailed schematic of the experimental setup is provided in Figure~\ref{ExpTheo}(a). To analyze the geometry of a hovering droplet, the part above the undisturbed surface $D_s$, indicated in Figure~\ref{ExpTheo}(b), and the droplet diameter $D$ were measured from a photograph taken when the droplet was at the working distance $d_w$ of the camera lens. The camera recorded 60 frames per second (FPS) with a macro lens (1:1 reproduction ratio). Each frame contains $1920\times1080$ pixels. To minimize the error involved in measuring $D_s$, the bath was filled slightly above the rim to achieve an almost eye-level camera angle. Calibration was done by placing a ruler in front of the camera at $d_w$. To capture the dynamics of droplet deposition and self-propulsion, droplets were also recorded with an iPhone~5s and a high-speed camera. Top view recording at 200~FPS allowed for measurement of the horizontal component $s_h$ of the displacement $\bm{s}$ as a function of time $t$. The respective horizontal components $v_h$ and $a_h$ of the velocity $\bm{v}$ and the acceleration $\bm{a}$ in the direction of $\bm{s}$ were obtained from $s_h$ and presented after using a Savitzky--Golay filter with a window of 9 data points (45~ms) and a polynomial order equal to unity.

To record fluid flow, high-speed camera imaging at 2500~FPS was performed. Spherical hollow micro balloons of approximate diameter 50~$\mu m$ (supplied by SIG Mfg.\ Co.\ Inc.) were used as tracer particles in an acetone droplet or on a liquid substrate. Outside the droplet, an aerosol was used to visualize the gas flow. The aerosol was produced by holding dry ice in the vicinity of the droplet, leading to the condensation of water from air. For these experiments, the droplets were also illuminated from the side or from above so that the camera, the bath, and the light source formed an angle of 90\degree.

Table~\ref{tbl:1} shows the lowest temperatures of substrates on which acetone droplets are deposited and subsequently hover in a LS. These values were obtained by depositing droplets as close to the substrates as possible and with $D \approx 2l_c$, where $l_c = \sqrt{\gamma_{13}/\rho_3 g} = 1.7$~mm denotes the capillary length of the droplet at $56\degree$C, with $\rho_{3}$ and $g$ being the mass density of the droplet and the gravitational acceleration on earth, respectively.

\section{Non-coalescence}
\label{Noncoalescence}

\begin{figure}
\begin{overpic}{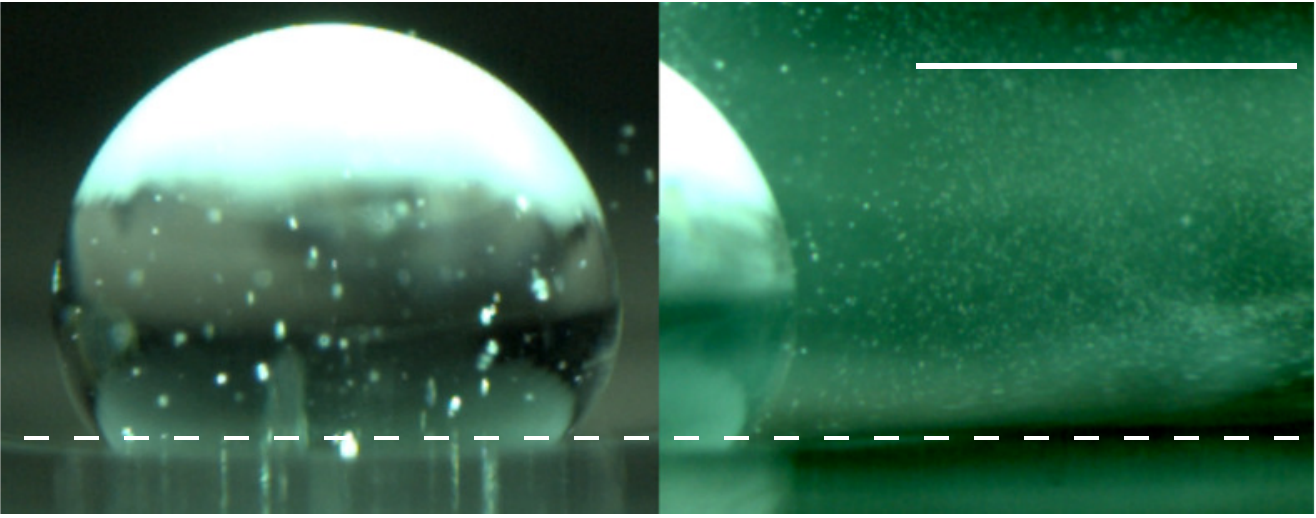}
\put(2,36){\color[rgb]{1,1,1}(a)}
\put(52,36){\color[rgb]{1,1,1}(b)}
\put(80,35.5){\color[rgb]{1,1,1}2 mm}
\end{overpic}
\caption{Acetone droplets ($D \approx 2.9$~mm) hovering on 75\degree C water. The pictures come from a sequence of images obtained at 2500~FPS. The dashed line indicates the position of the undisturbed water surface and the scale bar applies to both images. In (a) the micro balloon tracer particles are inside the acetone droplet and move in the center of the droplet from the top to the bottom with a velocity of about $20 \pm 5$~cm/s. (Multimedia view) In (b) the aerosol which is made from small water droplets and surrounds the droplet. By tracing the small water droplets it is possible to visualise the gas flow close to the acetone droplet and the liquid substrate. (Multimedia view)}
\label{DropletsTracer}
\end{figure}

Figure~\ref{DropletsTracer}(a) (Multimedia view) depicts an acetone droplet with tracer particles hovering on 75\degree C water. This image is from a sequence of images obtained at 2500~FPS. The associated movie is slowed down by a factor of 250. The majority of the tracer particles that are in focus when the edge in the middle of the droplet is in focus travel from the top to the bottom in nearly straight line trajectories. Due to lensing effects, only those particles which pass through the center of the droplet can be observed clearly. The tracer particles have a velocity of about $20 \pm5~\text{cm/s}$. When the front of the droplet is in focus, the particles move from the bottom to the top. The tracer particles have a velocity of about $15 \pm 5~\text{cm/s}$. When this motion is caused by a thermocapillary effect, the tracer particles follow the thermal characteristic Marangoni flow with speed $v_M$, which scales as
\begin{equation}
    \label{Eq1}
   v_M \sim \frac{\Delta\gamma}{\mu},
\end{equation}
where $\Delta \gamma$ is the magnitude of the difference in surface tension caused by a temperature difference $\Delta T$ over the droplet and $\mu$ is the dynamic viscosity of the liquid comprising the droplet. For acetone close to $T_{sat}$, $\mu = 0.2$~mPa\,s and for $v_M = 15$~cm/s, we have $\Delta \gamma = 30~\mu$N/m, which yields $\Delta T \approx 0.3$~K. Tokugawa and Takaki\cite{tokugawa_mechanism_1994} showed that a temperature difference in this order is typically obtained for a LD in which a Marangoni thermocapillary flow is present.

\begin{figure}
\begin{overpic}{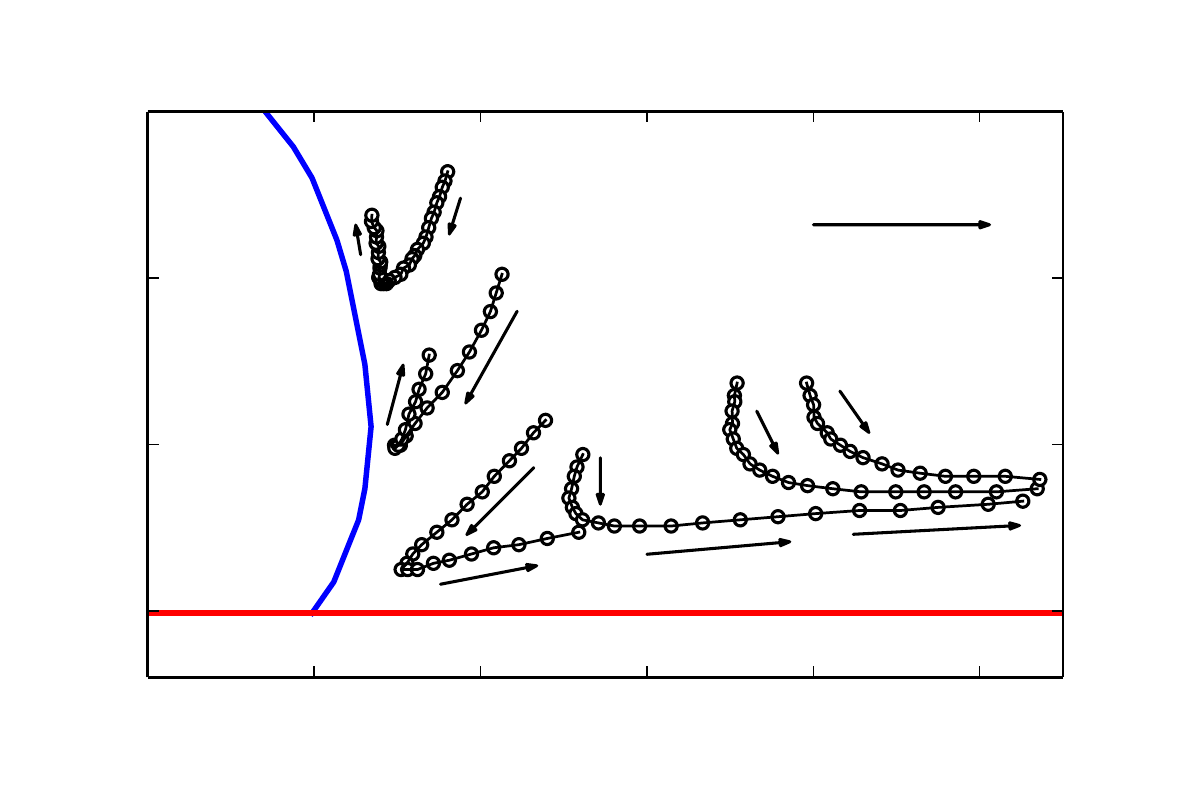}
    \put(10.5,5.886039){0.0}
    \put(25.0,5.886039){0.5}%
    \put(39.0,5.886039){\color[rgb]{0,0,0}\makebox(0,0)[lb]{\smash{1.0}}}%
    \put(53.0,5.886039){\color[rgb]{0,0,0}\makebox(0,0)[lb]{\smash{1.5}}}%
    \put(67.0,5.886039){\color[rgb]{0,0,0}\makebox(0,0)[lb]{\smash{2.0}}}%
    \put(81.5,5.886039){\color[rgb]{0,0,0}\makebox(0,0)[lb]{\smash{2.5}}}%
    \put(7.5,14.297198){\color[rgb]{0,0,0}\makebox(0,0)[lb]{\smash{0.0}}}%
    \put(7.5,28.388108){\color[rgb]{0,0,0}\makebox(0,0)[lb]{\smash{0.5}}}%
    \put(7.5,42.479017){\color[rgb]{0,0,0}\makebox(0,0)[lb]{\smash{1.0}}}%
    \put(7.5,56.569925){\color[rgb]{0,0,0}\makebox(0,0)[lb]{\smash{1.5}}}%
    \put(33.636364,50.524244){\color[rgb]{0,0,0}\makebox(0,0)[lb]{\smash{1}}}%
    \put(42.372727,46.015152){\color[rgb]{0,0,0}\makebox(0,0)[lb]{\smash{2}}}%
    \put(45.19091,33.333334){\color[rgb]{0,0,0}\makebox(0,0)[lb]{\smash{3}}}%
    \put(51.109091,29.951517){\color[rgb]{0,0,0}\makebox(0,0)[lb]{\smash{4}}}%
    \put(62.100001,36.433335){\color[rgb]{0,0,0}\makebox(0,0)[lb]{\smash{5}}}%
    \put(68.863637,36.715152){\color[rgb]{0,0,0}\makebox(0,0)[lb]{\smash{6}}}%
    \put(45,11.75){\color[rgb]{1,0,0}\makebox(0,0)[lb]{\smash{Liquid Substrate}}}%
    \put(15.730557,37.684497){\color[rgb]{0,0,1}\makebox(0,0)[lb]{\smash{Droplet}}}%
    \put(66.5,50){\color[rgb]{0,0,0}\makebox(0,0)[lb]{\smash{ $|\bm{v}_t| = 25$~cm/s}}}%
    \put(45,1.5){\color[rgb]{0,0,0}\makebox(0,0)[lb]{\smash{\textit{x}-axis (mm)}}}%
    \put(4,30){\color[rgb]{0,0,0}\rotatebox{90}{\makebox(0,0)[lb]{\smash{\textit{z}-axis (mm)}}}}%
\end{overpic}
\caption{Flow and velocity $\bm{v}_t$ of six small water droplets of an aerosol near the acetone droplet ($D \approx 2.9$~mm) depicted in Figure~\ref{DropletsTracer}(b). The speed $v_t = \left| \bm{v}_t\right|$ of the tracer particles scales with the length of the arrows.}
\label{FlowData}
\end{figure}

\begin{figure}
\begin{overpic}{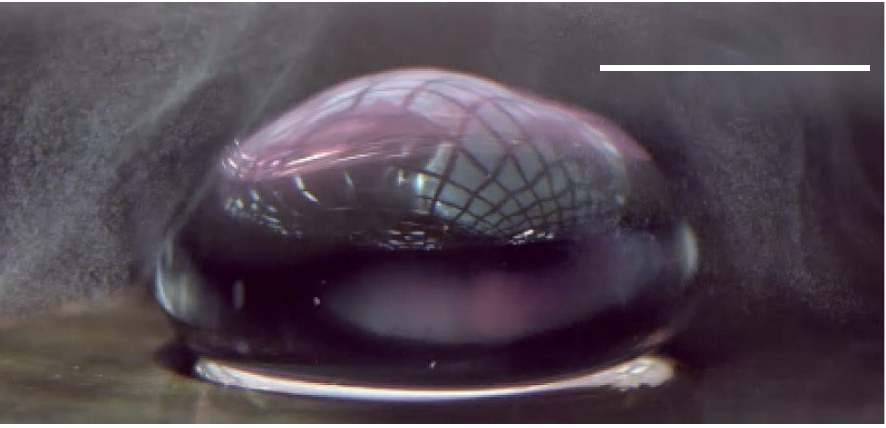}
\put(79,43){\color[rgb]{1,1,1}3 mm}
\end{overpic}
\caption{Acetone droplet ($D \approx 6$~mm) in a Leidenfrost state hovering on a polished copper substrate with a temperature of $\approx200$\degree C. The picture comes from a sequence of images obtained at 2500~FPS. (Mulitmedia view) An aerosol which is made from small water droplets surrounds the acetone droplet. The reflected pattern on the surface of the droplet is from a basket containing the dry ice which is used to generate the aerosol.}
\label{CopperSubstrate}
\end{figure}

\begin{figure}
\begin{overpic}{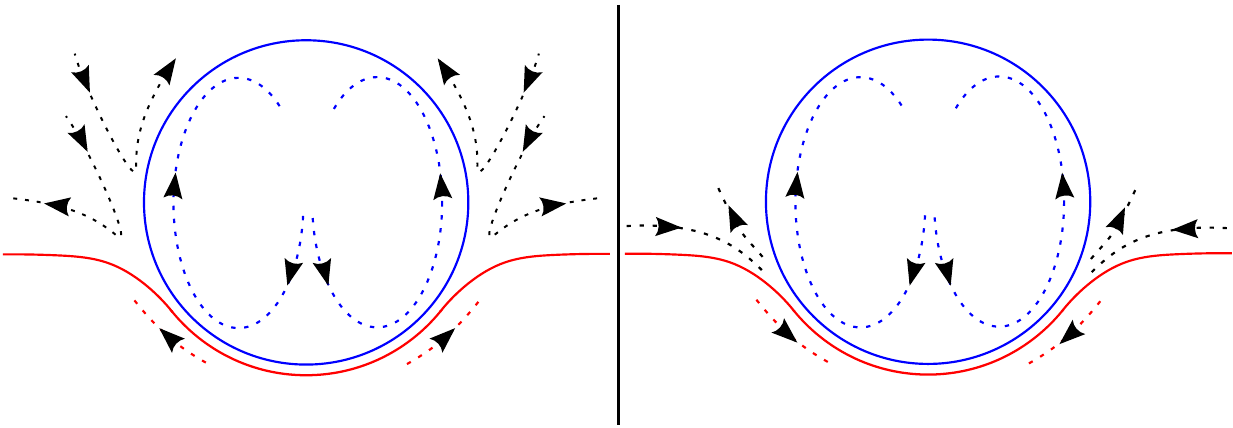}
\put(2,33){(a)}
\put(52,33){(b)}
\put(22.25,22.5){\color[rgb]{0,0,1}cold}
\put(20.5,19){\color[rgb]{0,0,1}droplet}
\put(72.75,22.5){\color[rgb]{0,0,1}cold}
\put(71,19){\color[rgb]{0,0,1}droplet}
\put(1,21){gas}
\put(51,21){gas}
\put(1,9){\color[rgb]{1,0,0}hot}
\put(1,5.5){\color[rgb]{1,0,0}liquid}
\put(1,2){\color[rgb]{1,0,0}substrate}
\put(51,9){\color[rgb]{1,0,0}hot}
\put(51,5.5){\color[rgb]{1,0,0}liquid}
\put(51,2){\color[rgb]{1,0,0}substrate}
\end{overpic}
\caption{(a) Schematic depicting the flows of gas and liquid generated by an acetone droplet hovering on 75\degree C water. The Marangoni solutocapillary flow on the substrate points away from the droplet and the gas flow points towards the gap between the droplet and the substrate. The extent to which the latter flow influences non-coalescence is unclear. (b) Schematic depicting the flows of gas and liquid generated by placing a cold droplet above a hot liquid substrate. The Marangoni thermocapillary flow of the substrate points towards the droplet and causes classical non-coalescence by self-lubrication.}
\label{FlowScheme}
\end{figure}

Figure~\ref{DropletsTracer}(b) (Multimedia view) depicts an acetone droplet surrounded by an aerosol. This image is from a sequence of images obtained at 2500~FPS. The associated movie is again slowed by a factor of 250. This aerosol makes it possible to visualize the gas flow close to the surfaces of the supporting layer and the droplet. Figure~\ref{FlowData} shows the streamlines and velocity $\bm{v}_t$ of six small water droplets of the aerosol imaged in Figure~\ref{DropletsTracer}(b), which are easily traced. Assuming that no-slip boundary conditions hold, the gas flow on the outside of the acetone droplet follows the expected poloidal flow in the droplet, as might be expected from the analysis of the tracer particles in the droplet. The gas close to the substrate is directed away from the evaporating acetone droplet. This could be due to gas emission from the gap between the droplet and the surface of the water substrate. Alternatively, it could stem from a Marangoni solutocapillary flow at the surface of the substrate. This flow would originate form the mixing of acetone vapour and water on the liquid surface. Under these circumstances, the surface tension of the liquid substrate close to the evaporating droplet would be lower than elsewhere in the bath and this would generate a flow which is directed away form the droplet along the surface of the substrate. 

Our observations show that floating particles in the vicinity of the droplet move strongly away from the droplet, indicating that a solutocapillary flow is indeed present. To investigate if the presence of acetone vapor can induce a solutocapillary Marangoni flow, a controlled experiment with an acetone droplet which hangs on the tip of an uncharged needle and with a water bath which is packed with micro balloons, both at room temperature, was conducted. The micro balloons cover about 10\% of the total surface of the liquid substrate and are almost evenly spaced. The micro balloons strongly move away form the evaporating acetone droplet, thereby opening a hole on the surface of about 2~cm in diameter. When the acetone droplet is replaced by a water droplet, no hole forms. If the liquid surface in the vicinity of a liquid droplet can have a surface tension value which is lower than the surface tension value of water, then only a small number of acetone molecules are needed to fully cover the surface of the liquid substrate. Since acetone and water are miscible, they mix in all proportions. After removing the acetone droplet, the micro balloons slowly close the previously created hole in a process which takes several seconds. This indicates that the adsorbed acetone molecules evaporate in air and/or mix with the liquid substrate. In subsection~\ref{Quasi-static acetone droplets on water}, additional evidence is provided to show that the surface tension of the liquid substrate strongly decreases in the vicinity of the evaporating acetone droplet.

In Figure~\ref{FlowData}, the flow of aerosol particle 3 indicates that a strong gas flow towards the gap between the droplet and the substrate is present. This flow seems to be mostly generated by the solutocapillary Marangoni flow. Gas emission from the gap between the droplet and the water substrate can be reasoned to have a small impact on the observed flows because we were not able to observe the strong gas flow towards and away from the gap between an acetone LD and a solid substrate. Figure~\ref{CopperSubstrate} (Multimedia view) shows an acetone LD ($D \approx 6$~mm) on a polished copper substrate with a temperature of $\approx 200\degree$C. The associated movie is slowed by a factor of 83. No clear signs of a horizontal flow generated by gas emission from the gap between the droplet and the copper substrate are present. The gas flow in the vicinity of the droplet seems to be mostly generated by the Marangoni flow in the droplet. Additionally, Le Merrer et al.\cite{le_merrer_wave_2011} reasoned that for a liquid nitrogen droplet hovering on water, which evaporates much more forcefully than the acetone droplets considered in this work, the emitted gas has an escape speed of 1~cm/s. This is about 30 times lower than the most rapid gas flow observed in this work. Finally, the schematic in Figure~\ref{FlowScheme}(a) summarizes the observations and shows that non-coalescence for acetone droplets hovering on a liquid substrate at a higher temperature is established in a different way than in Figure~\ref{FlowScheme}(b), which represents classical non-coalescence by self-lubrication. As mentioned above, Marangoni thermocapillary flows in liquid bodies, which originate from thermal gradients, can prevent coalescence. Savino et al.\cite{savino_marangoni_2003} showed that the surface of the liquid substrate moves towards the droplet and causes gas to flow towards the gap between the droplet and the liquid substrate, whereby coalescence is prevented due to the presence of a self-lubricating layer. It is also worth mentioning that in our work non-coalescence is only observed for $T_S > T_{sat}$. As explained in the introduction, this is a characteristic condition which must be met for an LD to exist. Savino et al.\cite{savino_marangoni_2003} 
showed that non-coalescence of droplets by thermocapillary effects, a consequence of which is self-lubrication, can be obtained for $T_S < T_{sat}$, with the condition that a temperature difference between the droplet and the surface exists.

With the obtained evidence, it is possible to conclude that non-coalescence is not caused by classical non-coalescence by self-lubrication but at least partly by the Leidenfrost effect. For this reason, some discussions in this work are given in the light of the LE. So far, the extent to which the strong air flow towards the gap between the droplet and the substrate influences non-coalescence is unclear.

\section{Statics}
\label{Statics}
\subsection{Theory}
Consider a droplet of diameter $D$ formed by a liquid of mass density $\rho_3$ hovering above a liquid substrate of mass density $\rho_2$, as depicted in Figure~\ref{ExpTheo}(b). If all effects associated with the surface tension of the liquid surface are neglected and the droplet is modeled as a rigid sphere that is immersed by a distance $D-D_s>0$, then the volume $V_s$ of the portion of the droplet above the undisturbed surface of the substrate can be expressed as
\begin{equation}
\label{Eq2}
V_s=\frac{\pi D^3}{6}\Big(1-\frac{\rho_3}{\rho_2}\Big).
\end{equation}
Equating (\ref{Eq2}) to the expression $\pi D_s^2(3D-D_s)/6$ for the volume of a spherical cap of height $D_s$ cut from a sphere of diameter $D>D_s$, we find that $D_s/D$ must satisfy the cubic equation
\begin{equation}
\label{Eq7}
\left(\frac{D_s}{D}\right)^3-\frac{3}{2}\left(\frac{D_s}{D}\right)^2
+\frac{1}{2}\left(1-\frac{\rho_3}{\rho_2}\right)=0.
\end{equation}
Granted that $\rho_3<\rho_2$, (\ref{Eq7}) can be solved analytically and gives only one physically acceptable root. In particular, for a rigid sphere with the mass density of acetone floating on water, we have $\rho_2(T = 70\degree \text{C})=970$~kg/m$^3$ and $\rho_3(T = T_{sat})\approx750$~kg/m$^3$ and (\ref{Eq2}) gives $D_s/D\approx0.31$.

If, alternatively, the surface tension $\gamma_{12}$ of the liquid substrate, which is assumed to be uniform, is taken into consideration but while continuing to model the droplet as a rigid sphere, it is still possible to determine $D_s/D$. To obtain $D_s$ for a given choice of $D$, we used an iterative numerical scheme based on the work of Rapacchietta and Neumann\cite{rapacchietta_force_1977} for approximating the depth $d_0$ and the angle $\psi$ of the effective contact line of a perfectly non-wetting sphere. Each iteration involves three steps. In the first step, we choose $\psi$ and with a fixed-point shooting method find $d_0$ by integrating the static axisymmetric Young--Laplace equation
\begin{equation}
\label{YL}
\gamma_{12}\left\{ \frac{z''}{\left(1+z'^2\right)^{3/2}} + \frac{z'}{r\left(1+z'^2\right)^{1/2}} \right\} = \left(\rho_{2}-\rho_{1}\right)gz,
\end{equation}
where $\rho_{1}$ denotes the mass density of the air and a prime denotes differentiation with respect to the radial coordinate $r$, subject to the respective near-field and far-field conditions
\begin{equation}
\label{BCs}
z'(r_0) = \tan{\beta(r_0)}
\qquad\text{and}\qquad
\lim_{r\to\infty}z'(r)\to0.
\end{equation}
{The geometrical relations $r_0 = (D/2)\sin{\psi}$ and $\beta(r_0) = \psi$ then provide the additional information needed to determine all  unknown quantities. Before integrating, (\ref{YL}) and (\ref{BCs}) are first parametrized with the angle $\beta$ to avoid numerical problems involving large values of $r$. In the second step, we evaluate the vertical component $F_z$ of the resultant force which acts on the droplet}, accounting for the weight $\bm{F}_g=-\rho_3g(\pi D^3/6)\bm{e}_z$ of the droplet, the upward component $\bm{F}_s=2\pi r_0\gamma_{12}(\bm{t}\cdot\bm{e}_z)\bm{e}_z=2\pi r_0\gamma_{12}(\sin\psi)\bm{e}_z$ of resultant force distributed on the edge of the spherical portion of the liquid surface, and the buoyancy
\begin{equation}
\label{Fb}
\bm{F}_b=\frac{\pi\left(\rho_{2}-\rho_{1}\right)gD^3}{8}\left(\frac{2}{3} - \cos{\psi} + \frac{1}{3} \cos^3{\psi}+\frac{2d_0}{D} \sin^2{\psi}\right)\bm{e}_z,
\end{equation}
of the droplet. In the third step, we use a fixed-point method to update the value of $\psi$ given the previously determined value of $F_z$, towards obtaining a vertical force balance. The three steps are repeated until a satisfactory value for $\rho_3$ is obtained $F_z/F_g = 0 \pm 10^{-3}$. In the final step, the geometrical relation $D_s = (D/2)(1+\cos{\psi})-d_0$ is used to determine $D_s/D$.

Viscous forces can be ignored since the capillary number of all flows is much smaller than unity. By calculating $\textit{We}$ for the flows we can estimate the impact of inertia. Due to the low mass density of the air, \textit{We} for the air flow is well below unity and inertia of the gas can be ignored. The inertia of the liquids is difficult to estimate. The characteristic length might be described as the thickness of a liquid layer close to the liquid--air interface which moves due to Marangoni flow. If we guess that the thickness of such a layer is on the order of 1~mm and has an average velocity of approximately 15~cm/s, $\textit{We} = 1$. This indicates that inertia might play a role. However, as a first approximation we ignore forces induced by hydrodynamic effects.

\subsection{Slowly propagating acetone droplets on water}
\label{Quasi-static acetone droplets on water}
\begin{figure*}
\begin{overpic}[scale=0.85]{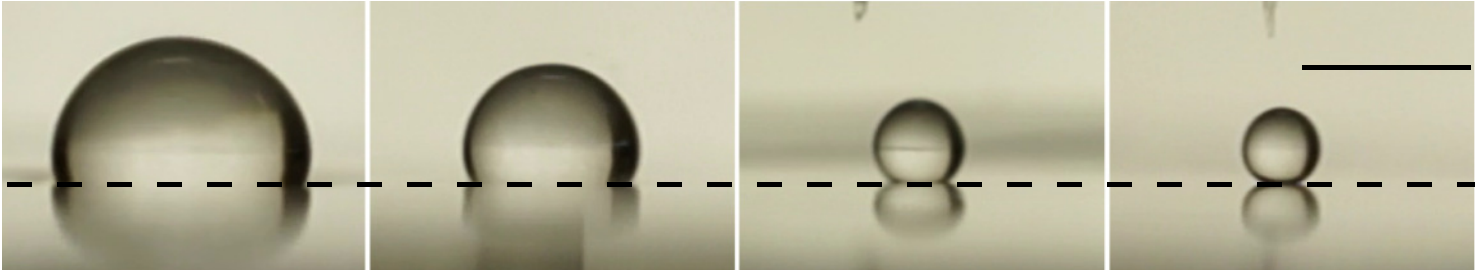}
\put(89.5,15){2.5 mm}
\end{overpic}
\caption{Acetone droplets in a Leidenfrost state on a $70\degree$C water bath. The dashed line indicates the position of the undisturbed water surface and the scale bar applies to all four images. (Multimedia view)}
\label{Droplets}
\end{figure*}
\begin{figure}
\begin{overpic}{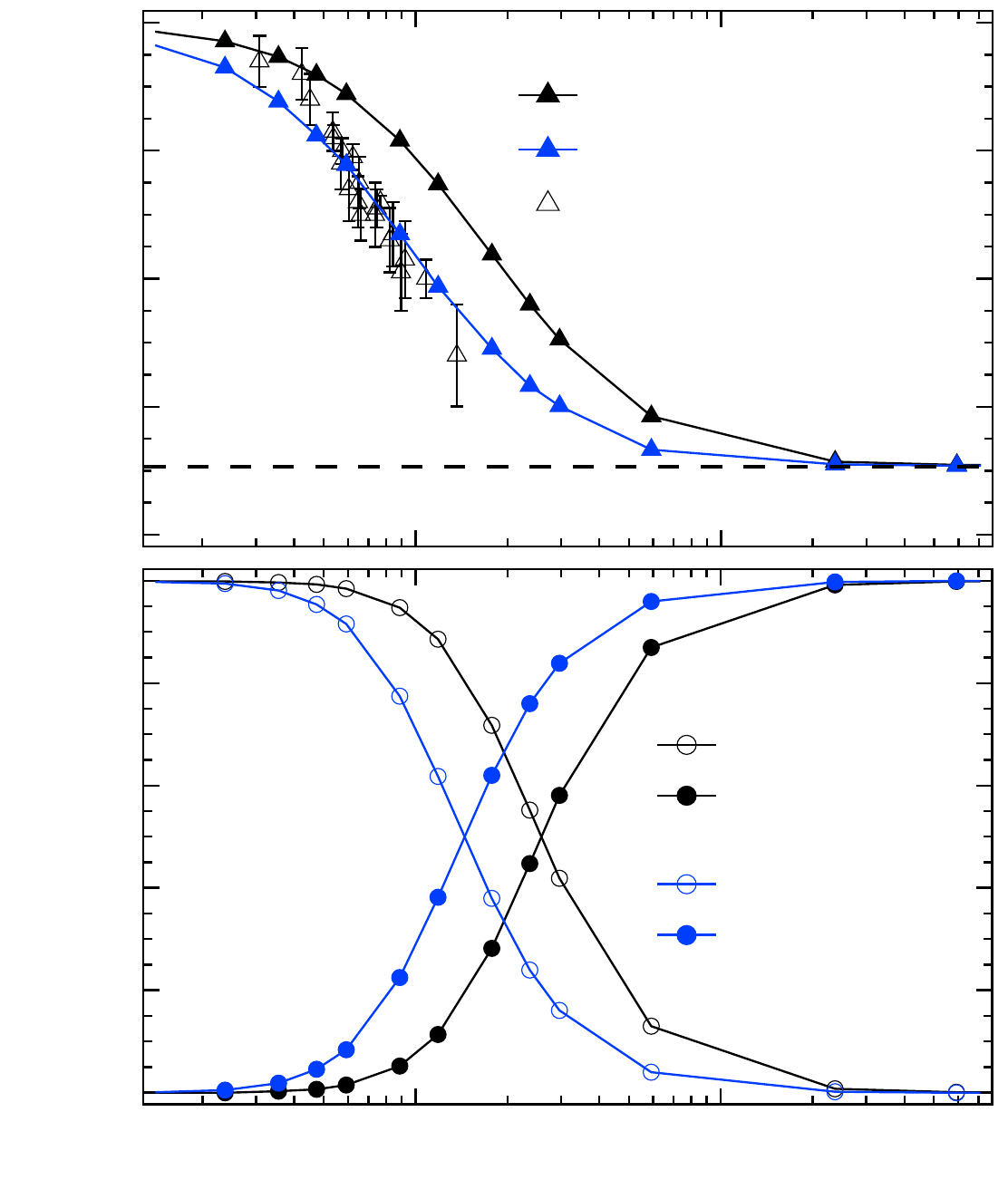}
    \put(7.5,97.5){\color[rgb]{0,0,0}\makebox(0,0)[lb]{\smash{1.0}}}%
    \put(7.5,86.7){\color[rgb]{0,0,0}\makebox(0,0)[lb]{\smash{0.8}}}%
    \put(7.5,76.1){\color[rgb]{0,0,0}\makebox(0,0)[lb]{\smash{0.6}}}%
    \put(7.5,65.5){\color[rgb]{0,0,0}\makebox(0,0)[lb]{\smash{0.4}}}%
    \put(7.5,54.8){\color[rgb]{0,0,0}\makebox(0,0)[lb]{\smash{0.2}}}%
    \put(49,91.5){\color[rgb]{0,0,0}\makebox(0,0)[lb]{\smash{Theory $\gamma_{12} = 64$~mN/m }}}%
    \put(49,86.9){\color[rgb]{0,0,0}\makebox(0,0)[lb]{\smash{Theory $\gamma_{12} = 20$~mN/m }}}%
    \put(49,82.4){\color[rgb]{0,0,0}\makebox(0,0)[lb]{\smash{Experiment}}}%
    \put(1,72){\rotatebox{90}{$D_s/D$}}%
    \put(1,27){\rotatebox{90}{$\text{Force} / \text{Weight}$}}%
    \put(7.5,51){\color[rgb]{0,0,0}\makebox(0,0)[lb]{\smash{1.0}}}%
    \put(7.5,42.3){\color[rgb]{0,0,0}\makebox(0,0)[lb]{\smash{0.8}}}%
    \put(7.5,33.8){\color[rgb]{0,0,0}\makebox(0,0)[lb]{\smash{0.6}}}%
    \put(7.5,25.3){\color[rgb]{0,0,0}\makebox(0,0)[lb]{\smash{0.4}}}%
    \put(7.5,16.9){\color[rgb]{0,0,0}\makebox(0,0)[lb]{\smash{0.2}}}%
    \put(7.5,8){\color[rgb]{0,0,0}\makebox(0,0)[lb]{\smash{0.0}}}%
    \put(33.5,4.5){\color[rgb]{0,0,0}\makebox(0,0)[lb]{\smash{$10^0$}}}%
    \put(59,4.5){\color[rgb]{0,0,0}\makebox(0,0)[lb]{\smash{$10^1$}}}%
    \put(55,41){\color[rgb]{0,0,0}\makebox(0,0)[lb]{\smash{$\gamma_{12} = 64$~mN/m }}}%
    \put(55,29){\color[rgb]{0,0,0}\makebox(0,0)[lb]{\smash{$\gamma_{12} = 20$~mN/m }}}%
    \put(61,37.2){\color[rgb]{0,0,0}\makebox(0,0)[lb]{\smash{$F_s/F_g$}}}%
    \put(61,33){\color[rgb]{0,0,0}\makebox(0,0)[lb]{\smash{$F_b/F_g$}}}%
    \put(61,25.5){\color[rgb]{0,0,0}\makebox(0,0)[lb]{\smash{$F_s/F_g$}}}%
    \put(61,21.3){\color[rgb]{0,0,0}\makebox(0,0)[lb]{\smash{$F_b/F_g$}}}%
    \put(45,0){\color[rgb]{0,0,0}\makebox(0,0)[lb]{\smash{$D/(2l_c)$}}}%
\end{overpic}
\newline
\caption{The unfilled triangles represent $D_s/D$ for experimental values of acetone droplets in a Leidenfrost state and the filled triangles are theoretical values for non-wetting spheres with the same mass density of acetone on a $70\degree$C liquid substrate. For $D/(2l_c) < 1$, with $l_c$ being the capillary length, droplets exhibit essentially spherical shapes. The horizontal dashed line is the theoretically predicated value of $D_s/D$ in the absence of surface tension. The unfilled circles and the filled circles show the respective theoretically predicted values of $F_s/F_g$ and $F_b/F_g$.}
\label{StaticData}
\end{figure}
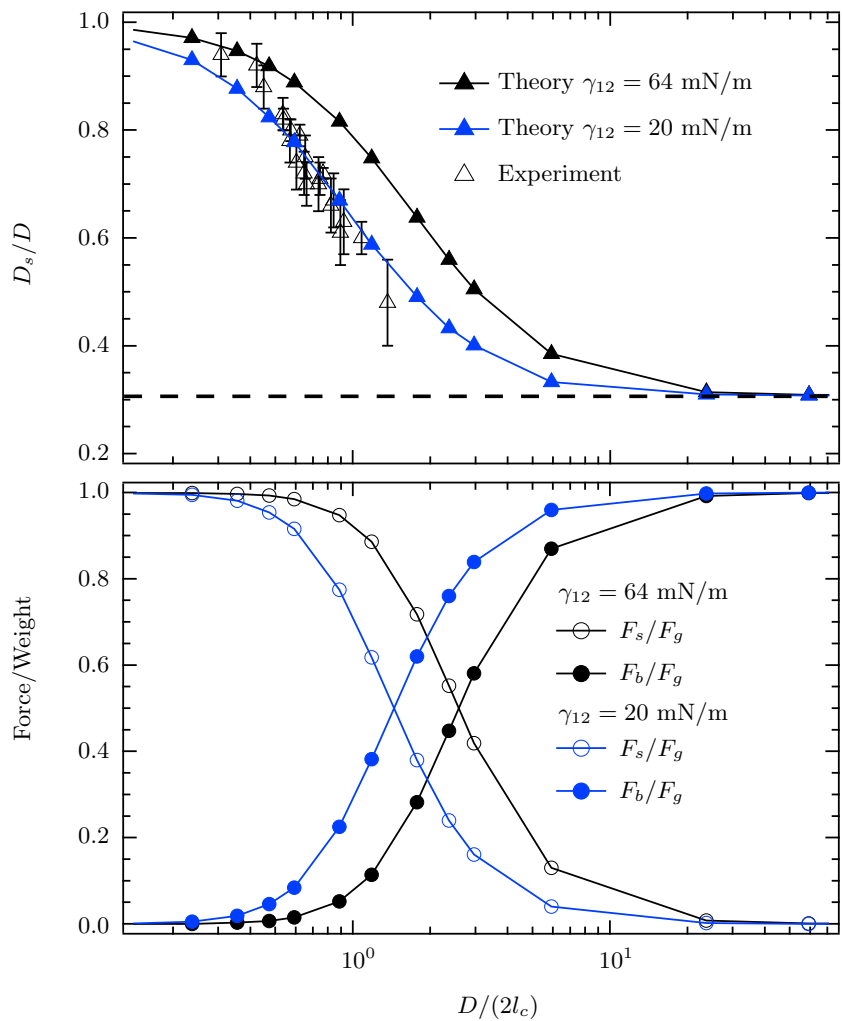
It is possible to measure $D_s$ immediately after deposition, when the velocity of an acetone droplet is low relative to the velocity that induces gravity-capillary waves on the surface of the liquid substrate. Figure~\ref{Droplets} (Multimedia view) shows a number of acetone droplets on a $70\degree$C water bath. It is evident that these droplets sink deeper beneath the undisturbed surface and become less spherical with increasing $D$. In Figure~\ref{StaticData}, the measured values of $D_s/D$ versus $D/(2l_c)$, where $l_c$ denotes the capillary length, are represented by the unfilled triangles. It is noteworthy that the ratio $D/(2l_c)$ is the square root of the Bond number $\textit{Bo}$, with characteristic length $D/2$. Mahadevan and Pomeau\cite{mahadevan_rolling_1999} and Aussillous and Qu{\'e}r{\'e}\cite{aussillous_properties_2006} showed that on a solid substrate, two situations may arise: for $D/(2l_c) < 1$, droplets exhibit essentially spherical shapes; for $D/(2l_c) > 1$, droplets form puddles. For a droplet on a liquid substrate and $D/(2l_c) > 1$, sphericity is strongly affected. This is expressed by the error incurred by measuring $D_s/D$ for the droplet with the largest diameter $D$. Values for larger droplets are omitted since their shapes deviate significantly from spherical. Recently, Ooi et al.\cite{ooi_deformation_2015} used a model to predict the geometry of super-hydrophobic droplets, in the form of liquid marbles, on a liquid substrate. However, since that model underestimates $D_s/D$ for $D/(2l_c)$ below unity, it is not useful in the present context. For $D/(2l_c)$ greater than unity, when the sphericity of the droplet is strongly affected, that model might be useful. The filled triangles represent the numerical values of $D_s/D$ for perfectly non-wetting spheres, with the mass density of acetone, on a $70\degree$C liquid substrate with the mass density of water. The black triangles are for a liquid substrate which only contains water ($\gamma_{12} = 64$~mN/m) and the blue triangles are for a liquid substrate which contains a high concentration of acetone ($\gamma_{12} \approx 20$~mN/m). When a surface tension gradient is taken into consideration, we assume that the value of $D_s/D$ falls between those for the black and blue triangles. The result of a force analysis in which an acetone droplet is modeled as a non-wetting perfect sphere provided in Figure~\ref{StaticData} provides an approximation for the force needed to support a droplet in the LS. The unfilled circles stand for $F_s/F_g$ and the filled circles stand for $F_b/F_g$. The analysis shows that, in our experiments, $F_b$ is dominated by $F_s$. The horizontal line in Figure~\ref{StaticData} at 31\% is $D_s/D$ if $\gamma_{12}$ is ignored and, as expected, for large values of $D/(2l_c)$ the numerical values of $D/D_s$, which are provided for completeness, tend toward this value.

Figure~\ref{StaticData} shows evidence indicating that the surface tension of the liquid substrate in the vicinity of the droplet is close to that of acetone and, thus, below the surface tension of water. The distance around the droplet which most strongly influences $D/D_s$ is $l_c$. Since this distance is approximately 2~mm, acetone can induce a strong effect on $D/D_s$ without migrating significantly. The mixing of acetone with water is expected as a consequence of Raoult's law, which dictates that if acetone vapor is present above water it should mix with water. This additionally suggests the presence of a strong Marangoni solutocapillary flow on the liquid substrate. The steeper slope of the experimental data points compared to the theory can be attributed to low acetone vapor production of small droplets and to deviation from spherical of large droplets. A low acetone vapor production can lead to values of $\gamma_{12}$ approaching 64~mN/m. Because our theory neglects the flattening of the droplets present in experiments, we overestimate the values of $D/D_s$.

\subsection{Effective Surface Tension}

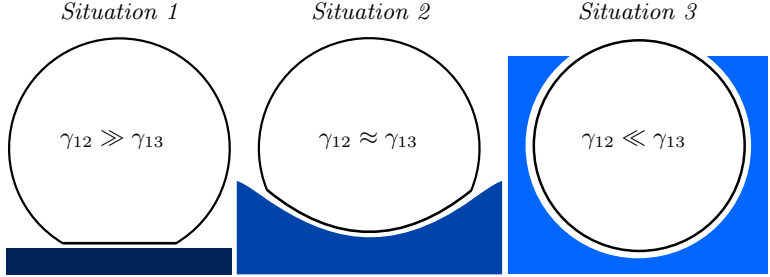
\begin{figure}
\begin{overpic}{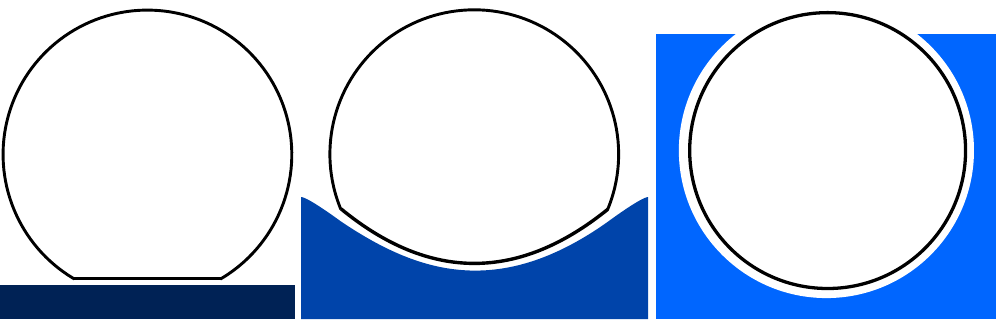}
    \put(7,34){\color[rgb]{0,0,0}\makebox(0,0)[lb]{\smash{\textit{Situation 1}}}}%
    \put(40,34){\color[rgb]{0,0,0}\makebox(0,0)[lb]{\smash{\textit{Situation 2}}}}%
    \put(75,34){\color[rgb]{0,0,0}\makebox(0,0)[lb]{\smash{\textit{Situation 3}}}}%
    \put(7,17.5){\color[rgb]{0,0,0}\makebox(0,0)[lb]{\smash{$\gamma_{12} \gg \gamma_{13}$}}}%
    \put(41,17.5){\color[rgb]{0,0,0}\makebox(0,0)[lb]{\smash{$\gamma_{12} \approx \gamma_{13}$}}}%
    \put(75.5,17.5){\color[rgb]{0,0,0}\makebox(0,0)[lb]{\smash{$\gamma_{12} \ll \gamma_{13}$}}}%
\end{overpic}
\caption{Schematic representations of droplets with surface tension value $\gamma_{13}$ on substrates with surface tension value $\gamma_{12}$.
}
\label{EffectiveST}
\end{figure}

Focusing on situations in which $D/(2l_c) < 1$, we next propose a simple way to tentatively evaluate the thickness $e$ of the vapor layer for a LD hovering above a liquid substrate. We begin by considering the pressure $P$ exerted by a droplet on a liquid surface. The surface and the droplet are separated by a thin gaseous layer and do not interact across the associated gap. In the event that $\gamma_{12} \gg \gamma_{13}$, as depicted in \textit{Situation~1} of Figure~\ref{EffectiveST}, pressure $P$ can be approximated by the pressure which is exerted by a LD on a flat and rigid surface. For this situation, Biance et al.\cite{biance_leidenfrost_2003} reasoned that $P$ can be approximated with the Laplace pressure $P_L$ for a sphere 
\begin{equation}
    \label{LP}
    P_L = \frac{4\gamma_{13}}{D}.
\end{equation}
For the opposite situation $\gamma_{12} \ll \gamma_{13}$ depicted in \textit{Situation~3} of Figure~\ref{EffectiveST}, $P$ is generated solely by buoyancy. Due to the small value of the associated lift force per unit area relative to the Laplace pressure, we may assume that $P \approx 0$ and neglect deviations from sphericity. For the intermediate situation $\gamma_{12} \approx \gamma_{13}$ depicted in \textit{Situation~2} of Figure~\ref{EffectiveST}, a value for $P$ between those of \textit{Situations~1} and \textit{3} should be obtained. Also, the sphericity of the droplet should be less significant than in \textit{Situation~1} but more significant than in \textit{Situation~3}. To evaluate $P$ in conjunction with (\ref{LP}), we introduce the effective surface tension $\gamma$, which depends on $\gamma_{12}$ and $\gamma_{13}$. As a first approximation the deformation in the vertical direction of the liquid surface and the droplet scales with the vertical component of the force acting on the droplet and the surface. Hence, it is reasonable to represent a droplet and a substrate by two springs in series. The effective spring constant for springs in series can be written as $1/k = 1/k_s + 1/k_d$ with $k_s$ and $k_d$ the respective spring constants of the substrate and the droplet. If we then assume that $k_s = C \gamma_{12}$ and $k_d = C \gamma_{13}$, with $C>0$ being a scale factor, a simple expression for $\gamma$ results
\begin{equation}
    \label{ES}
    \frac{1}{\gamma} = \frac{1}{\gamma_{12}} + \frac{1}{\gamma_{13}}.
\end{equation}
\emph{Situations 1, 2, and 3} correspond respectively, to choosing $\gamma_{12}\gg\gamma_{13}$ (so that $\gamma\approx\gamma_{13}$), $\gamma_{12}\approx\gamma_{13}$ (so that $\gamma\approx\gamma_{13}/2$), and $\gamma_{12}\ll\gamma_{13}$ (so that $\gamma\approx\gamma_{12}$). Using $\gamma$ in (\ref{LP}) shows that $P$ is reduced by a factor two for \textit{Situation~2} compared to \textit{Situation~1} and confirms that (\ref{ES}) is a potential candidate to describe $\gamma$.

Biance et al.\cite{biance_leidenfrost_2003} showed that the rate of evaporation $J$ through the surface of a LD obeys scaling relations $J\propto A/e$, with area $A$ of the droplet in close contact to the substrate, and $J\propto e^3 P$, from which we deduce that $e\propto\sqrt[4]{A/P}$. Since $P$ is inversely proportional to $A$, we infer that $e\propto 1/\sqrt{P}$ and, with reference to (\ref{LP}) and (\ref{ES}), obtain
\begin{equation}
\label{erel}
e\propto \sqrt{\frac{D}{\gamma}}.
\end{equation}
A simple consequence of (\ref{erel}) is that $e$ increases as $\gamma$ diminishes for fixed values of $D$. On this basis, it is possible to conclude that the value of $e$ for a liquid substrate is larger than that for a solid substrate. This occurs simply because liquid substrates are more compliant than hard solid substrates. This suggests that the presence of a thicker vapor layer makes a droplet hovering on a liquid substrate more stable than one hovering on a solid substrate. In our experiments, a radially outward Marangoni flow can nevertheless pump gas out of the vapor layer, thereby reducing the magnitude of $e$.

Antonini et al.\cite{antonini_water_2013} showed that when the evaporation rate across a surface is large enough, levitation can be enhanced. Therefore, it is essential to note that a water substrate at a temperature $T_S = 70\degree$C at which most of our experiments are performed has a saturated vapor pressure of approximately 0.3 atm. As a first approximation we can scale the saturated vapor pressure of a liquid with its evaporation rate. Assuming that the acetone droplet has a temperature of 30$\degree$C during impact and the liquid substrate (pure water) a temperature of $70\degree$C, the ratio of water evaporation rate over acetone evaporation rate is $\approx  0.7$. This shows that evaporation of the liquid substrate can enhance levitation and plays a significant role in our experiments.

\section{Dynamics}
\subsection{Deposition}
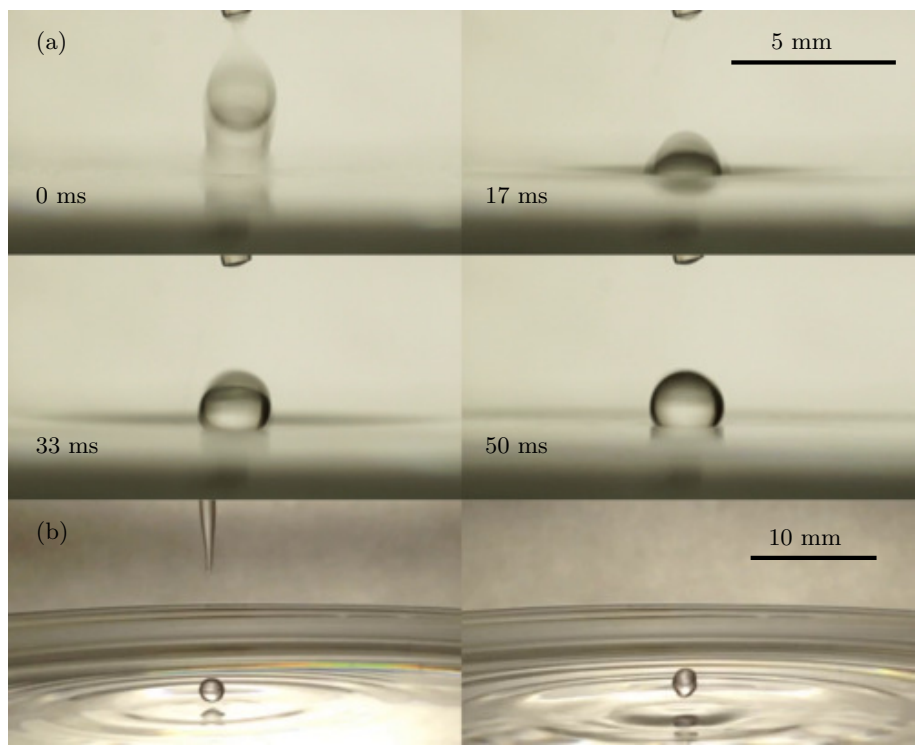
\begin{figure}
\begin{overpic}{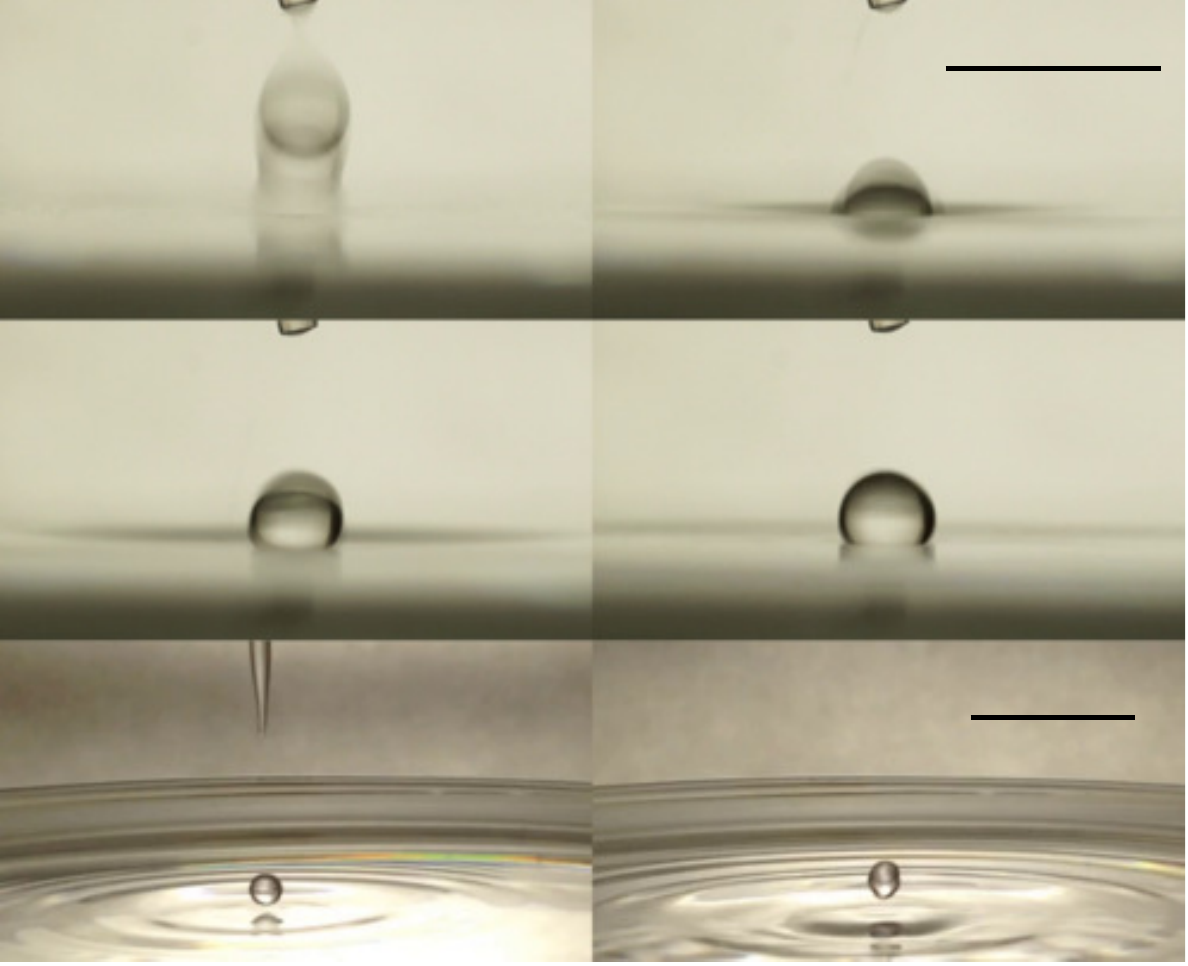}
\put(3,77){(a)}
\put(84,77){5 mm}
\put(3,60){0 ms}
\put(52.5,60){17 ms}
\put(3,32.5){33 ms}
\put(52.5,32.5){50 ms}
\put(3,23){(b)}
\put(83.7,22.3){10 mm}
\end{overpic}
\caption{Deposition of acetone droplets ($D = 2$~mm) on a $\text{70--75}\degree$C water bath. In (a), a droplet is deposited with a relatively low impact velocity. The scale bar applies to all four images presented in (a). The water substrate deforms heavily under impact and the droplet undergoes a damped oscillation. On the left hand side of (b), the droplet bounces cleanly off the substrate after impacting it with a speed of $45\pm5$~cm/s ($\textit{We} \approx 14 $). On the right hand side of (b), the droplet bounces cleanly off the substrate for an impact speed of $70\pm5$~cm/s ($\textit{We} \approx 35 $). For higher impacting velocities the droplets coalesce with the water bath. The scale bar applies to both images presented in (b).}
\label{DropletDepos}
\end{figure}

A sequence of images depicting the impact of a 2~mm diameter acetone droplet on $70\degree$C water is shown in Figure~\ref{DropletDepos}(a). This droplet is at room temperature before it absorbs heat from the substrate. The droplet rebounds after initially sinking and experiences a damped oscillation without losing contact with the substrate. This behavior was also observed by Lee and Kim,\cite{lee_impact_2008} who   described two additional regimes. In the second of the these regimes, which occurs when the impacting speed $v_i$ of a sphere exceeds a certain value, the sphere bounces cleanly off the substrate. To reach the third regime, $v_i$ must be large enough so that the sphere sinks. The boundaries of the three regimes are determined by a dimensionless relationship of the form
\begin{equation}
    \label{LK}
    \textit{We}\mskip1.5mu\textit{Bo}^{3/2} = n\bigg(\frac{\rho_2}{\rho_3}\bigg)^{\!\!2},
\end{equation}
where $n>0$ is a prefactor and the Bond and Weber numbers $\textit{Bo}$ and $\textit{We}$ are defined according to 
\begin{equation}
\textit{Bo} =\frac{ \rho_2 g D^2 }{\gamma_{12}}
\qquad\text{and}\qquad
\textit{We} = \frac{\rho_{2} v_i^{2} D}{\gamma_{12}}.
\end{equation}
Lee and Kim scale $v_i$ by $1/\rho_3$, which explains the appearance of $\rho_3$ in the denominator on the right-hand side of (\ref{LK}). To distinguish the three regimes, two values of $n$ need to be measured. By increasing $v_i$, we also observed bounce-off. In Figure~\ref{DropletDepos}(b), acetone droplets ($D = 2$~mm) with $v_i = 45\pm 5$~cm/s (left) and $v_i = 70\pm5$~cm/s (right) bounce cleanly off $75\degree$C water. The corresponding values of $\textit{We}$ are approximately equal to 14 and 35, respectively. We also noticed that for $v_i$ between 45--65~cm/s, the chance of successfully depositing droplets with a diameter of 2~mm was lower than for values ranging between $\text{0--50}$~cm/s and between $\text{65--70}$~cm/s. For a series of experiments with $D = \text{1.6--2.8}$~mm, we observed bounce-off for all droplets if $v_i = \text{65--70}$~cm/s. For this range of $v_i$, larger droplets tend to disintegrate, which is evident from the formation of two droplets, one significantly smaller than the other. A summary of these findings appears in Table~\ref{tbl:2}. It is clear that for larger droplets a forbidden regime for $v_i$ exists, namely $v_i = \text{45--65}$~cm/s. In that regime, droplets coalesce with the substrate.
The mechanisms underlying these observations remain uncertain. For future reference, a potentially useful resource concerning impact-induced coalescence is the work of Planchette et al.,\cite{planchette_coalescence_2013} who used a geometric criterion to describe the transition between non-coalescence and coalescence of droplets with armored interfaces.
\begin{table}[!t]
  \caption{The behavior of acetone droplets after impacting a 75\degree C water substrate as a function of impact speed and diameter. The errors on these quantities are $\pm 0.05$~mm and $\pm 5$~cm/s respectively. A = damped oscillation; B = bouncing; C = disintegration; and D = coalescence.}
  \label{tbl:2}
  \begin{tabular}{lcccc}
    \hline
    $v_i$ (cm/s) & $\text{0--45}$ & $\text{45--65}$ & $\text{65--70}$ & $>$70\\
    \hline
    $D$ (mm) & & & & \\
    $1.6-2.0$ & A & B & B & D\\
    $2.0-2.4$ & A & B/D & B & D\\
    $2.4-2.8$ & A & D & B/C & D\\
    \hline
  \end{tabular}
\end{table}
It is important to note that liquid droplets bouncing on solid substrates can exhibit remarkable elasticity, as Richard and Qu{\'e}r{\'e}\cite{richard_bouncing_2000} showed, which possibly enhances bounce-off in our experiments.

As reported by Tran et al.,\cite{tran_drop_2012} the dynamic Leidenfrost temperature $T_{Ld}$ rises steeply as $\textit{We}$ increases. This might stem from higher values of $P$ during impact. This potential explanation is supported by the experiments conducted by Baumeister et al.\cite{baumeister_metastable_1966} and Ng et al.,\cite{ng_suppression_2015} who reported that substrate vibrations are detrimental for the LS. Also, Qu{\'e}r{\'e}\cite{quere_leidenfrost_2013} noticed that the dynamic pressure of a droplet exceeds the value of $P$ for $\mathit{We} > 1$ and thereby emphasized that dynamic levitation is more difficult to achieve than static levitation. Consistent with the discussion in the statics section, the enhanced compliance of liquid substrates leads to lower values of $P$. As depicted in Figure~\ref{DropletDepos}(a), the shape of a liquid surface changes significantly during droplet impact. We therefore assume that values of $P$ in our experiments are lower than those in experiments on hard solid substrates. This might explain why such low values of $T_s$ can accommodate acetone droplets in a LS.
\subsection{Self-propulsion and Drag}
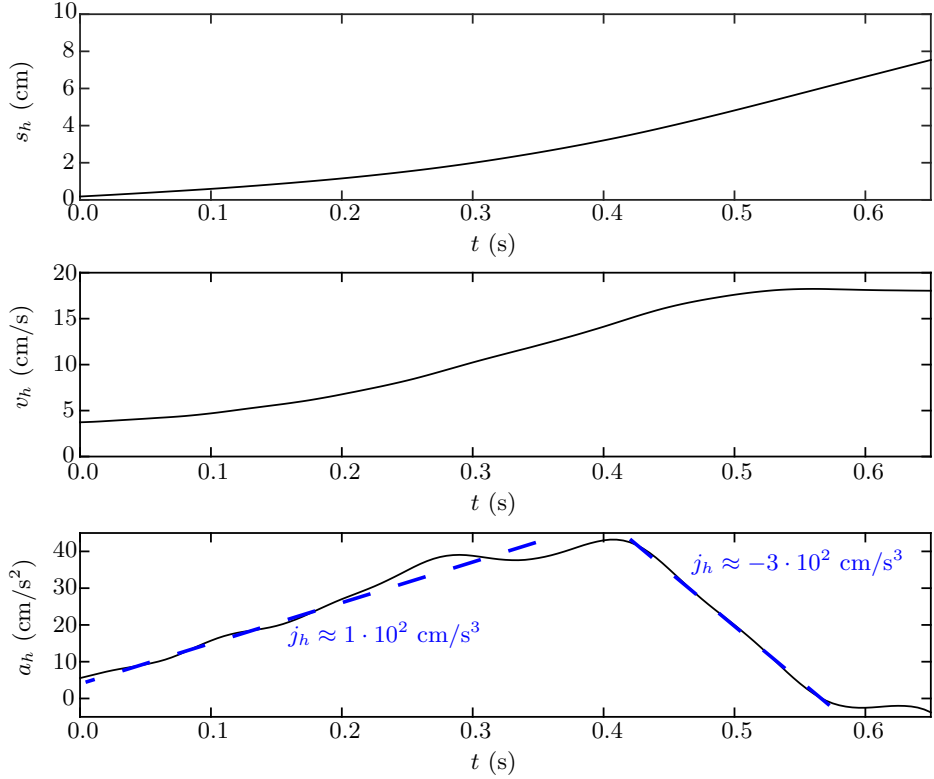
\begin{figure}
\begin{overpic}[scale=0.85]{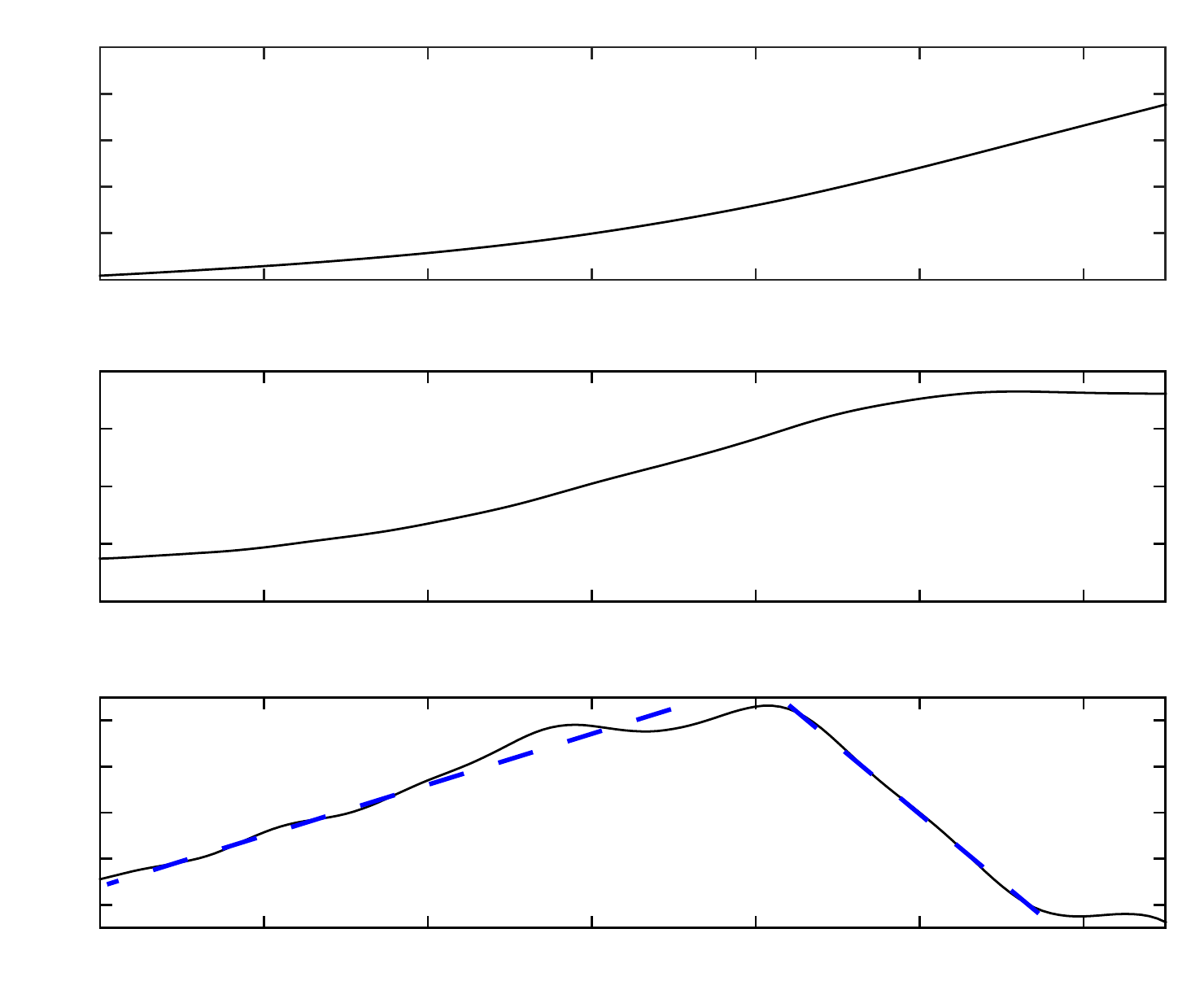}
	\put(49,0.75){$t$ (s)}
\put(7,4){0.0}
\put(20.6,4){0.1}
\put(34.3,4){0.2}
\put(47.9,4){0.3}
\put(61.6,4){0.4}
\put(75.1,4){0.5}
\put(88.8,4){0.6}
    	\put(1,10.5){\rotatebox{90}{$a_h$ (cm/s$^2$)}}%
\put(6.5,7.5){0}
\put(5.5,11.25){10}
\put(5.5,15.1){20}
\put(5.5,19){30}
\put(5.5,22.9){40}
\put(30,14){{\clb $j_h \approx 1\cdot10^2$ cm/s$^3$}}
\put(72,21.5){{\clb $j_h \approx -3 \cdot10^2$ cm/s$^3$}}
	\put(49,27.75){$t$ (s)}
\put(7,31){0.0}
\put(20.6,31){0.1}
\put(34.3,31){0.2}
\put(47.9,31){0.3}
\put(61.6,31){0.4}
\put(75.1,31){0.5}
\put(88.8,31){0.6}
    	\put(1,38.25){\rotatebox{90}{$v_h$ (cm/s)}}%
\put(6.5,32.75){0}
\put(6.5,37.5){5}
\put(5.5,42.25){10}
\put(5.5,47){15}
\put(5.5,51.75){20}
\put(49,54.75){$t$ (s)}
\put(7,58){0.0}
\put(20.6,58){0.1}
\put(34.3,58){0.2}
\put(47.9,58){0.3}
\put(61.6,58){0.4}
\put(75.1,58){0.5}
\put(88.8,58){0.6}
    	\put(1,66){\rotatebox{90}{$s_h$ (cm)}}%
\put(6.5,59.5){0}
\put(6.5,63.25){2}
\put(6.5,67.1){4}
\put(6.5,71){6}
\put(6.5,75){8}
\put(5.5,79){10}
\end{overpic}
\caption{The respective horizontal components $s_h$, $v_h$, and $a_h$ of the displacement $\bm{s}$, the velocity $\bm{v}$, and the acceleration $\bm{a}$ of a $3.0 \pm 0.1$~mm diameter droplet on a $79 \pm 1\degree$C water bath in the direction of $\bm{s}$ and as a function of time $t$. The slope of each dashed line represents the horizontal component $j_h$ of the time derivative of $a_h$ and is obtained with simple linear fitting.}
\label{DropletMotion}
\end{figure}
After formation, we observed that hovering droplets exhibit self-propulsion, accelerating in rectilinear trajectories. The direction in which a droplet travels can be imposed by imparting it with an initial velocity with a chosen magnitude and orientation. It is noteworthy that an LD can be deflected by the wall of the bath. Figure~\ref{DropletMotion} shows the respective horizontal components $s_h$, $v_h$, and $a_h$ of the displacement $\bm{s}$, the velocity $\bm{v}$, and the acceleration $\bm{a}$ of a $3.0 \pm 0.1$~mm diameter droplet on a $79 \pm 1 \degree$C water bath in the direction of $\bm{s}$ and as a function of time $t$. The droplet accelerates and reaches a maximum and essentially constant speed of $\approx$ 18~cm/s. For $t$ between 0--0.4~s, the horizontal component $j_h$ of the time derivative of $a_h$ is $\approx 1 \cdot 10^2$~cm/s$^3$. For $t$ between 0.4--0.6~s, $j_h \approx -3 \cdot 10^2$~cm/s$^3$, and for $t > 0.6$~s, $j_h$ and $a_h$ are $\approx 0$. The values of $j_h$ are obtained with simple linear fitting. From all acetone droplets observed in this work ($D < 2l_c$), which are found to hover on water with substrate temperatures between 70--80\degree C, we typically obtain curves as Figure~\ref{DropletMotion} shows. No significant change of $D$ is observed during the experiments, from which we assume that droplets do not lose mass. In what follows, we give provisional explanations for our observations.
\begin{figure}
\begin{overpic}{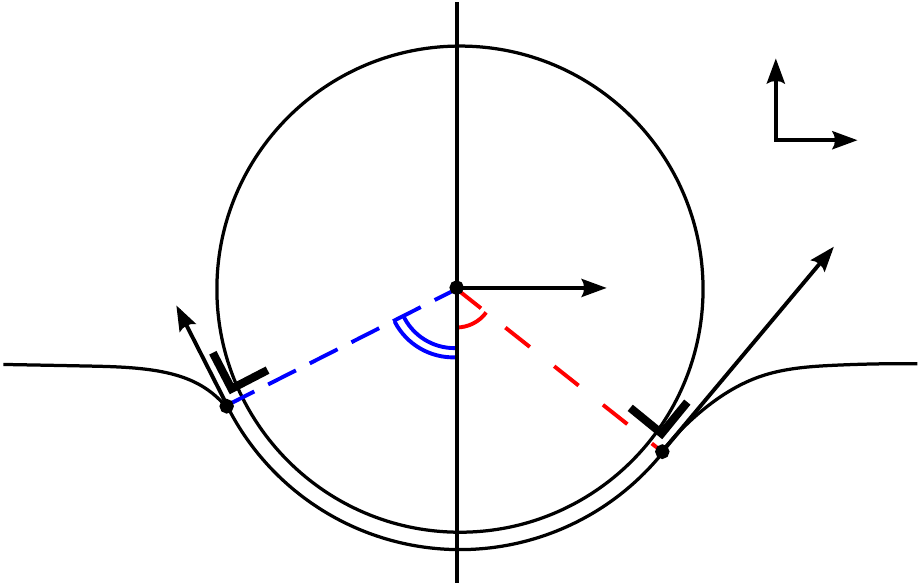}
    \put(90,45.125){\color[rgb]{0,0,0}\makebox(0,0)[lb]{\smash{$\bm{e}_{x}$}}}%
    \put(79.25,55){\color[rgb]{0,0,0}\makebox(0,0)[lb]{\smash{$\bm{e}_{z}$}}}%
    \put(10,38){\color[rgb]{0,0,0}\makebox(0,0)[lb]{\smash{1}}}%
    \put(10,10){\color[rgb]{0,0,0}\makebox(0,0)[lb]{\smash{2}}}%
    \put(40,38){\color[rgb]{0,0,0}\makebox(0,0)[lb]{\smash{3}}}%
    \put(60,35){\color[rgb]{0,0,0}\makebox(0,0)[lb]{\smash{$\bm{F}_{M}$}}}%
    \put(51,25){\color[rgb]{1,0,0}\makebox(0,0)[lb]{\smash{$\psi_f$}}}%
    \put(42,23){\color[rgb]{0,0,1}\makebox(0,0)[lb]{\smash{$\psi_b$}}}%
    \put(82,38){\color[rgb]{0,0,0}\makebox(0,0)[lb]{\smash{$\gamma_f \bm{t}_f$}}}%
    \put(11.5,27.5){\color[rgb]{0,0,0}\makebox(0,0)[lb]{\smash{$\gamma_b \bm{t}_b$}}}%
    \put(4,55){\color[rgb]{0,0,0}\makebox(0,0)[lb]{\smash{$\gamma_b < \gamma_f$}}}%
    \put(4,49){\color[rgb]{0,0,0}\makebox(0,0)[lb]{\smash{$\psi_b > \psi_f$}}}%
\end{overpic}
\caption{Schematic of a non-coalescing droplet is represented by a non-wetting solid cylinder. The surface tension $\gamma_f$ at the leading edge of the droplet is lower than that, $\gamma_b$, at the trailing edge of the droplet. The vectors $\bm{t}_f$ and $\bm{t}_b$ are of unit length and are tangent to the surface of the liquid substrate. The orthonormal basis vectors $\bm{e}_x$ and $\bm{e}_{z}$ correspond to the axes of a rectangular Cartesian coordinate system. In the scenario envisioned, a resultant force $\bm{F}_{M}$, measured per unit length cylinder, propels the cylinder perpendicular to its axis in the direction of $\bm{e}_x$. Definitions of the remaining symbols appear in the caption of Figure~\ref{ExpTheo}.}
\label{StaticModel}
\end{figure}
\label{Drag}
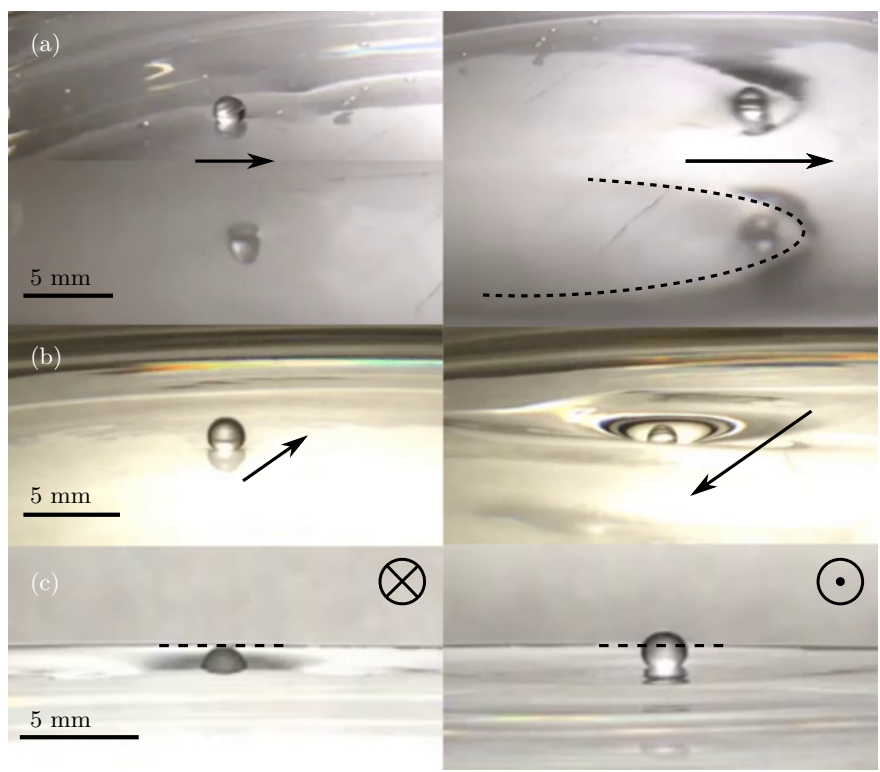
\begin{figure}
\begin{overpic}{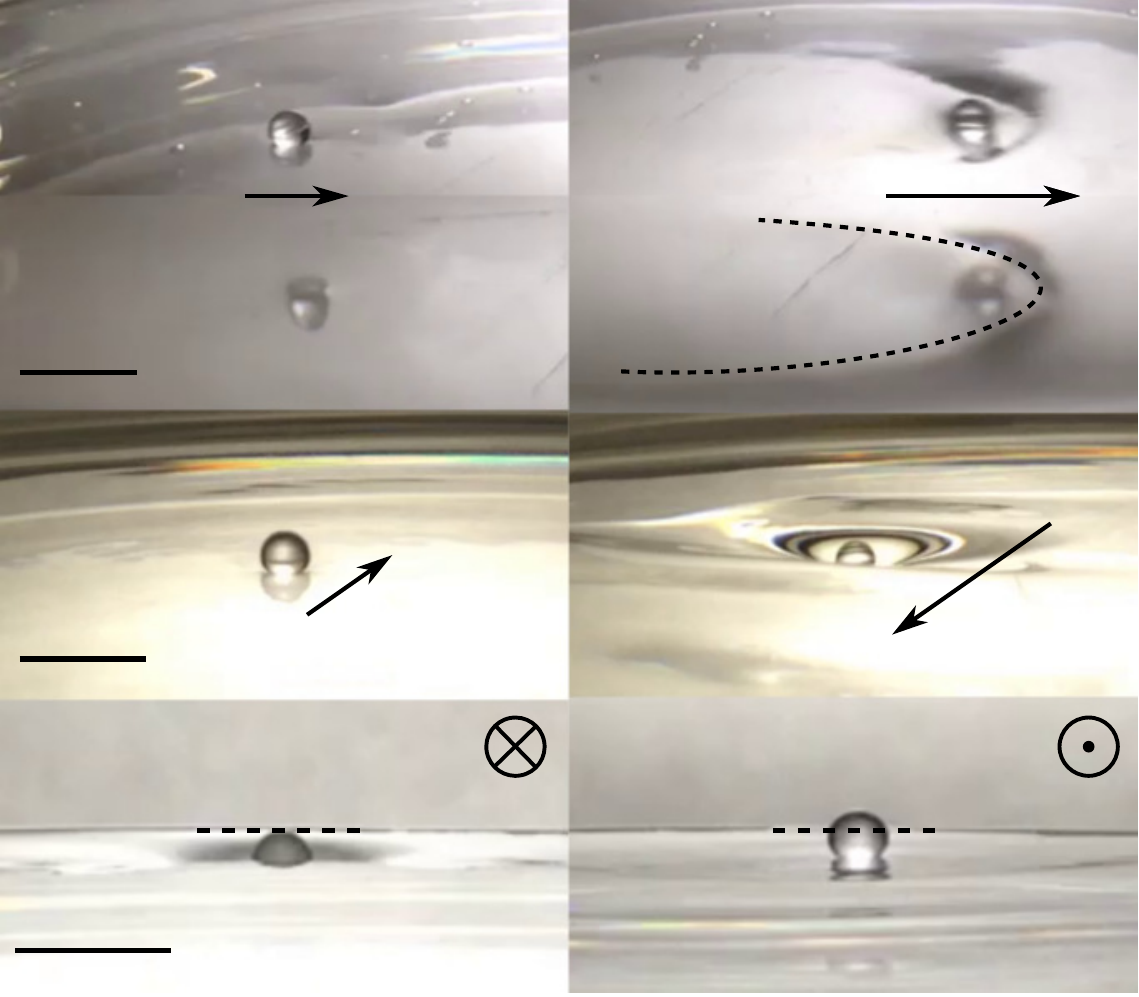}
\put(2.5,83){\color[rgb]{1,1,1}(a)}
\put(2.5,56){5 mm}
\put(2.5,47){\color[rgb]{1,1,1}(b)}
\put(2.5,31){5 mm}
\put(2.5,21){\color[rgb]{1,1,1}(c)}
\put(2.5,5.2){5 mm}
\end{overpic}
\caption{Acetone droplets in a Leidenfrost state ($D = 2$~mm) on a $75\degree$C water bath. The arrows indicate the direction of the velocity of the droplets. The scale bars apply to the left-hand and the right-hand frames. In (a), the reflection on the bottom of the bath is depicted together with the droplet. On the left-hand side of (a) the velocity of the droplet is relatively low and on the right-hand side of (a) the velocity of the droplet is relatively fast. The formation of a large feature in the vicinity of the droplet is clearly evident. On the left hand side of (b) a droplet with a velocity close to zero, which accelerates due to a Marangoni effect, is depicted. On the right-hand side of (b), the droplet passes the same location as on the left-hand side of (b) after being deflected by the wall of the water bath and approaches a maximum velocity of $19\pm1$~cm/s. The surface of the water bath deforms in a vicinity of the droplet. On the left hand side of (c), a rapidly propagating droplet reaches the wall of the water bath. Immediately after the droplet is deflected the velocity of the droplet is strongly reduced and the droplet becomes less immersed, as shown on the right-hand side of (c).}
\label{Propulsion}
\end{figure}

Linke et al.\cite{linke_self-propelled_2006} and Dupeux et al.\cite{dupeux_self-propelling_2013} showed that under certain conditions objects in a LS are known to be self-propelled. Burton et al.\cite{burton_geometry_2012} demonstrated that the thickness of the lubricating vapor layer of an LD on a flat substrate may exhibit non-axisymmetric features. Snezhko et al.\cite{snezhko_pulsating-gliding_2008} speculated that this effect can lead to self-propulsion. They observed ``ballistic'' flights, with fixed directions of less than 2~cm with a maximum $v_h$ of less than 2~cm/s, of small liquid nitrogen LDs on low viscosity liquids and assumed that the directional instability of the vapor layer abruptly changes the direction of propagation. Since the values of $v_h$ observed in our experiments are an order of magnitude larger and changes in direction are not seen, another mechanism appears to be operative. Recently, Bormashenko et al.\cite{bormashenko_self-propulsion_2015} observed self-propelled alcohol containing liquid marbles (70\% alcohol in water) on water, moving in rectilinear trajectories. Propulsion was reasoned to be a consequence of a Marangoni effect. The effect can be induced by the condensation of alcohol, evaporated from the liquid marble on the water substrate. This causes a surface tension gradient proximate to the droplet, manifested by a lower surface tension behind the droplet than in front of the moving droplet. This model was experimentally confirmed by Oshima et al.,\cite{oshima_surface_2014} who measured the surface tension in the front of and behind self-propelled hexanol droplets on an aqueous phase. It is important to bear in mind that hexanol is only slightly soluble in water and thus that a hexanol droplet does not have to be in the LS to float on a substrate. Bormashenko et al.\ assumed that $v_h$ is equal to the solutocapillary characteristic Marangoni flow which can be expressed as $\Delta \gamma/\mu$, where $\Delta \gamma$ is the magnitude of a surface tension difference and $\mu$ is the viscosity of the liquid substrate. The highest value of $v_h$ measured for the liquid marbles is approximately 15~cm/s. The lower values of $v_h$ found in the experiments of Bormashenko et al.\ can be expected since the acetone droplets evaporate more, which makes $\Delta \gamma$ increase. Additionally, granted that the droplets used in the experiments reported here were in the LS, this might lead to a significant reduction of drag. 

To explain our observations, we considered a two-dimensional model of which a schematic is depicted in Figure~\ref{StaticModel}. This model yields a horizontal resultant force $\bm{F}_M$ on an infinitely long cylinder (and measured per unit length of that cylinder). The sum 
\begin{equation}
    \label{sumgamma}
     \gamma_f \cos{\psi_f} - \gamma_b \cos{\psi_b}
\end{equation}
of $\gamma_f \bm{t}_f \cdot \bm{e}_{x} = \gamma_f \cos{\psi_f}$ and $\gamma_b \bm{t}_b \cdot \bm{e}_{x} = - \gamma_b \cos{\psi_b}$
generates this force. Moreover, $-\bm{F}_{M}=(\gamma_b \cos{\psi_b}-\gamma_f \cos{\psi_f})\bm{e}_{x}$ can be interpreted as the force that would be needed to fix the horizontal position of the cylinder. Since the surface of a liquid substrate with a lower surface tension bends more easily than the surface of a liquid substrate with a higher surface tension, we assume that $\psi_b > \psi_f$ if $\gamma_b < \gamma_f$. Therefore, $F_M = \bm{F}_M \cdot \bm{e}_{x}$ of $\bm{F}_M$ should be positive, which partly explains how a surface tension imbalance can cause a hovering object to move on a quiescent liquid substrate. With the mass of the droplet taken as constant, by neglecting water drag, and by assuming that $\psi_f \approx \psi_b$, we then obtain that $a_h \propto \Delta \gamma$. Figure~10 shows for $t$ between 0--0.4~s, that also $a_h \propto v_h$, assuming that the amount of evaporated acetone is constant in time. Then, we obtain that $\Delta \gamma \propto v_h$. So far, a simple model for predicting this relation is to the best of our knowledge unknown.

We now investigate whether the negative value of $j_h$ for $t$ between 0.4--0.6~s in Figure~\ref{DropletMotion} can be caused by wave drag. Rapha{\"e}l and de Gennes\cite{raphael_capillary_1996} derived that wave drag steeply increases when dimensionless number $\alpha = v_h/c_{min}$, with $c_{min} = (4g\gamma_{12}/\rho_2)^{1/4}$, reaches unity. They did this with a dispersion relation that capillary-gravity waves travel at minimum velocity $c_{min}$ so that a stationary wake of capillary-gravity waves forms for $\alpha > 1$. These waves carry energy away from the moving droplet, resulting in wave drag. In our experiments, $c_{min} \leq 22$~cm/s, which for $v_h =  19\pm1$~cm/s yields $\alpha = 0.9$ and $\alpha = 1.1$ for pure water and pure acetone substrates, respectively. The left-hand image in Figure~\ref{Propulsion}(a) shows a relatively slowly moving droplet and its reflection at the bottom of the water bath. For a relatively rapidly moving droplet in the image on the right of the same figure, the reflection clearly shows a large wave-like feature around the droplet. The conventionally expected wave radiation wedge does not seem to form, which indicates that in our experiments $\alpha$ remains slightly less than unity. Since the region in which the deflection of the surface of the substrate noticeable extends far around the droplet, it is reasonable to infer that the regions with low acetone concentrations are curved. Therefore, $\alpha$ can take a value slightly smaller than unity. We note that the value of $\alpha$ is most probably affected by a Marangoni flow. However, we can conclude that the negative value of $j_h$ can not be explained by wave drag according to the model of Rapha{\"e}l and de Gennes\cite{raphael_capillary_1996} since drag clearly appears for values of $\alpha$ below the treshold $\alpha = 1$ at which wave drag becomes important.

\begin{figure*}
\begin{overpic}[scale=0.85]{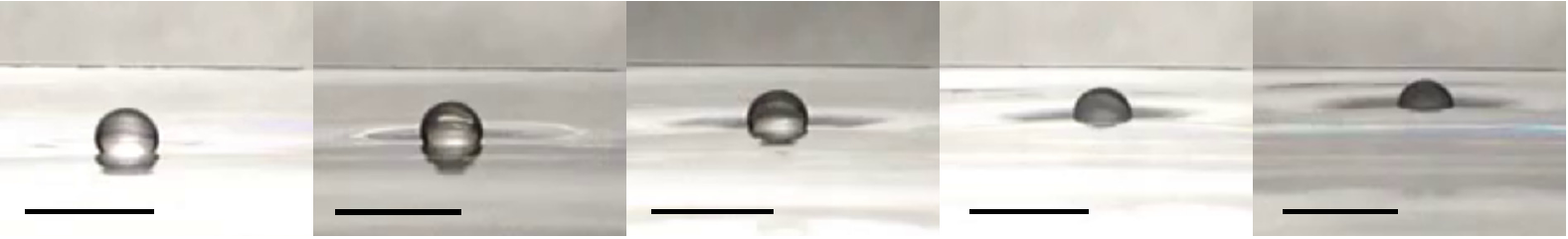}
    \put(2,12.5){\color[rgb]{0,0,0}\makebox(0,0)[lb]{\smash{0.40 s}}}%
    \put(22,12.5){\color[rgb]{0,0,0}\makebox(0,0)[lb]{\smash{0.43 s}}}%
    \put(42,12.5){\color[rgb]{0,0,0}\makebox(0,0)[lb]{\smash{0.47 s}}}%
    \put(62,12.5){\color[rgb]{0,0,0}\makebox(0,0)[lb]{\smash{0.50 s}}}%
    \put(82.5,12.5){\color[rgb]{0,0,0}\makebox(0,0)[lb]{\smash{0.53 s}}}%
    \put(1.7,2.4){\color[rgb]{0,0,0}\makebox(0,0)[lb]{\smash{4 mm}}}%
    \put(21.5,2.4){\color[rgb]{0,0,0}\makebox(0,0)[lb]{\smash{4 mm}}}%
    \put(41.5,2.4){\color[rgb]{0,0,0}\makebox(0,0)[lb]{\smash{4 mm}}}%
    \put(61.5,2.4){\color[rgb]{0,0,0}\makebox(0,0)[lb]{\smash{4 mm}}}%
    \put(82,2.4){\color[rgb]{0,0,0}\makebox(0,0)[lb]{\smash{4 mm}}}%
\end{overpic}
\caption{A self-propelled acetone droplet ($D = 2$~mm) which gradually becomes immersed in a $75\degree$C water bath and which moves away from the camera. At 0.4~s, the droplet has a horizontal velocity of $\approx 14$~cm/s and that velocity increases as a function of time. The change in light intensity is caused by flickering of the light box.}
\label{Sinking}
\end{figure*}
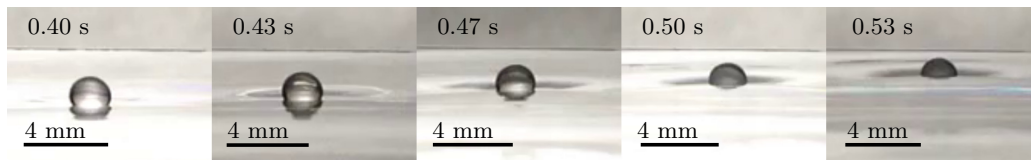

We further investigate the behavior of the droplets to find the phenomenon which underlies the negative values of $j_h$ that arise for $t$ between 0.4--0.6~s, as indicated in Figure~\ref{DropletMotion}. In the left-hand image of Figure~\ref{Propulsion}(b), a hovering acetone droplet with $\alpha$ close to zero accelerates due to a Marangoni effect. After bouncing against the wall of the water bath, the droplet accelerates in the opposite direction and reaches $v_h = 19\pm1$~cm/s, as depicted in the left-hand image of Figure~\ref{Propulsion}(b). In both images of Figure~\ref{Propulsion}(b), the droplet is located at approximately the same position but a strong deflection of the surface is clearly observable in the right-hand image. In the left-hand image of Figure~\ref{Propulsion}(c), a droplet reaches the wall of the water bath with $v_h = 19\pm1$~cm/s. After deflection, $\alpha$ decreases significantly and, as is evident from the dashed guidelines, the droplet is positioned distinctly higher relative to the undisturbed surface than previously. This shows that droplets that propagate sufficiently rapidly become immersed beneath the undisturbed water surface. These experiments were verified for a droplet in the middle of the water bath with an experimental setup involving two cameras: one to trace the in-plane position of the droplet and the other to detect the height of the droplet. Figure~\ref{Sinking} illustrates the gradually immersion of an acetone LD as a function of the time after deposition. This demonstrates that a moving pressure field induced by a non-wetting sphere is accompanied by the gradual formation of a depression under the undisturbed water surface. As a consequence, it is not surprising that the energy induced by a Marangoni effect is converted to an energy necessary to maintain the depression so that $F_M$ is balanced by the drag $F_d$ due to the formation of a depression. Therefore, the gradual formation of the depression for increasing values of $v_h$ should lead to a negative value of $j_h$.

Apart from water drag, $F_d$ might contribute to the drag on self-propelled liquid marbles, since the possibility exists that $F_d$ is present at $v_h = 15$~cm/s, which is approximately the maximum speed observed in the work of Bormashenko and coworkers.\cite{bormashenko_self-propulsion_2015}

Diorio et al.\cite{diorio_gravity-capillary_2009} showed that air flow deforms the surface. Moreover, the associated deflection deepens as $\alpha$ increases and for $\alpha > 1$, the conventionally expected wave radiation wedge forms. The deepening deflection as a function of $\alpha$ corresponds to the immersion of the acetone droplet in our work. 

As experiments reported by Le Merrer et al.\cite{le_merrer_wave_2011} show, wave drag steeply increases for decelerating liquid nitrogen LDs on a water substrate when $\alpha$ lowers and reaches unity. They use the model of Rapha{\"e}l and de Gennes\cite{raphael_capillary_1996} to explain their experimental observations. Still, drag is clearly observed for values of $\alpha$ below unity, for which no explanation is given. On the basis of the work presented here, we infer that this drag might be caused by the formation of a depression.

Burghelea and Steinberg\cite{burghelea_wave_2002} experimentally investigated drag on an immerserd rigid sphere and observed a gradual decrease in drag as a function of horizontal velocity, after subtracting viscous drag. The lowest drag was recorded at $\alpha \approx 0.9$. For higher values of $\alpha$, drag strongly increased due to the formation of gravity-capillary waves. In contrast to their work, the distance $d_{i}$ between the moving object and the undisturbed surface of a liquid substrate is not fixed during the experiments reported here. Therefore, to theoretically describe or experimentally measure drag on a small sphere or other small moving objects supported by a liquid substrate, it is crucial to leave $d_i$ free.

\section{Conclusions}
This exploratory work shows that non-coalescent droplets of acetone can be formed on liquid substrates. Remarkably, droplet deposition, subsequent hovering, and self-propulsion appear at low substrate temperatures and evidence for non-coalescence by the Leidenfrost effect is present. Despite the low temperature of the liquid substrate, it is found that the non-coalescence of acetone droplets is not a consequence of classical self-lubrication. Still, strong air flows which are generated by Marangoni flows are observed and their effect on non-coalescence remains unclear.

A case study suggests that small acetone droplets can be described by a simple model for a non-wetting rigid sphere hovering on a liquid substrate. To qualitatively evaluate the thickness of the vapor layer separating a small droplet and a liquid substrate, a notion of effective surface tension is introduced. On this basis, it is found that the vapor layer on a liquid substrate should be larger than on a solid substrate. This is attributed to the greater compliance of the liquid substrate and partly explains why droplets can be formed on liquid substrates with low temperatures, even at relatively high impact velocities. By carefully observing these impacting droplets, it is possible to record bounce-off from the substrate.

Self-propulsion of droplets in straight line trajectories can be attributed to a solutocapillary Marangoni effect. When self-propelled droplets accelerate, they gradually become immersed. This is reasoned to cause drag at velocities below those needed to induce wave drag.

\section{Acknowledgements}
\begin{sloppypar}
The authors thank the Japanese Society for the Promotion of Science for making this work possible.
Support from the Okinawa Institute of Science and Technology Graduate University with subsidy funding from the Cabinet Office, Government of Japan, is also gratefully acknowledged.
\end{sloppypar}

\end{document}